\documentclass[prb,twocolumn,amsmath,amssymb,eqsecnum]{revtex4-1}

\usepackage{hyperref}

\usepackage{graphicx}
\usepackage{dcolumn}
\usepackage{bm}
\usepackage{tikz}
\usepackage{subfigure}

\usepackage{braket}
\usepackage{natbib}
\usepackage{mathrsfs}
\usepackage{verbatim}
\usepackage[titletoc,toc]{appendix}

\begin{document}

\title{Chern-Simons theory of the magnetization plateaus of the spin-1/2 quantum XXZ Heisenberg model on Kagome Lattice}


\author{Krishna Kumar$^{1}$}
\author{Kai Sun$^{2}$}
\author{Eduardo Fradkin$^{1}$}
\affiliation{$^{1}$Department of Physics and Institute for Condensed Matter Theory, University of Illinois at Urbana-Champaign, 1110 West Green Street, Urbana, IL 61801}
\affiliation{$^{2}$Department of Physics, Randall Laboratory, University of Michigan, Ann-Arbor, Michigan 48109, USA}

\date{\today}

\begin{abstract}
Frustrated spin systems on Kagome lattices have long been considered to be a promising candidate for realizing exotic spin liquid phases. Recently, there has been a lot of renewed interest in these systems with the discovery of materials such as Volborthite and Herbertsmithite that have Kagome like structures. In the presence of an external magnetic field, these frustrated systems can give rise to magnetization plateaus of which the plateau at $m=\frac{1}{3}$ is considered to be the most prominent.  
Here we study the problem of the antiferromagnetic spin-1/2 quantum XXZ Heisenberg model on a Kagome lattice by using a Jordan-Wigner transformation that maps the spins onto a problem of fermions coupled to a Chern-Simons gauge field. This mapping relies on being able to define a consistent Chern-Simons term on the lattice. Such a lattice Chern-Simons term had previously only been written down for the square lattice and was used to successfully study the unfrustrated Heisenberg antiferromagnet on the square lattice. At a mean-field level, these ideas have also been applied to frustrated systems by ignoring the details of the Chern-Simons term. However, fluctuations are generally strong in these models and are expected to affect the mean-field physics. Using a recently developed method to rigorously extend the Chern-Simons term to the frustrated Kagome lattice we can now formalize the Jordan-Wigner transformation on the Kagome lattice. We then discuss the possible phases that can arise at the mean-field level from this mapping and focus specifically on the case of $\frac{1}{3}$-filling ($m=\frac{1}{3}$ plateau) and analyze the effects of fluctuations in our theory.  We show that in the regime of $XY$ anisotropy the ground  state at the $1/3$ plateau is equivalent to a bosonic fractional quantum Hall Laughlin state with filling fraction $1/2$ and that at the $5/9$ plateau it is equivalent to the first bosonic Jain daughter state at filling fraction $2/3$.
\end{abstract}

\maketitle


\section{Introduction}

Kagome lattice spin systems have been a topic of intense research for quite some time. It is believed that the high level of frustration in these systems can give rise to exotic spin liquid phases. A good model to look for these phases is the nearest-neighbor Heisenberg antiferromagnet on the Kagome lattice. In the past, many theoretical and numerical methods have been used to study such frustrated spin systems. More recently, with the discovery of materials like Volborthite and Herbertsmithite there is  the possibility of realizing some of these phases in actual experiments. 

The ground state of the quantum Heisenberg antiferromagnet on the Kagome lattice is considered to be a promising candidate for realizing spin liquid phases. However, so far the many theoretical and numerical studies still remain inconclusive. Recent numerical and theoretical studies strongly agree in favor of a gapped $\mathbb{Z}_2$ spin liquid state.\cite{Yan2011,Punk2014}. Other studies argue instead in favor of a gapless $U(1)$-Dirac spin liquid state.\cite{Ran2007,Iqbal2011} A few other studies also indicate the possibility of a valence-bond (VBC) type of crystalline state,\cite{Sachdev1992, Marston1991, Evenbly2010} while some other studies find symmetry breaking states \cite{Clark2013} or even a chiral spin liquid state.\cite{Yang1993,Messio2012} Further, recent experiments on Herbertsmithite do indicate that its ground state may indeed be a quantum spin liquid.\cite{Han2012}

In the presence of an external magnetic field, frustrated quantum antiferromagnets  are expected to give rise to magnetization plateaus of which the plateau at $m = \frac{1}{3}$ should be most prominent and, for this reason, there has been a lot of work analyzing the properties of these plateaus. Numerical works focusing on the isotropic and Ising regimes of the quantum Heisenberg antiferromagnet in an external magnetic field do identify magnetization plateaus at many different values. However, in the Ising regime, the simulations favor a valence bond crystal (VBC) type state with an enlarged unit cell based on a $\sqrt{3} \times \sqrt{3}$ structure at these plateaus.\cite{Cabra2005,Capponi2013,Nishimoto2013}

Experimentally it is difficult to observe these magnetization plateaus in Herbertsmithite (the most structurally perfect Kagome compound) since its exchange coupling is quite high, $J \approx 170 K$, which implies that it would require fields close to 200 T to be able to observe the $\frac{1}{3}$ plateau. As a result Okamoto {\it et al.} looked at Volborthite (with $J \approx 77 K$) and Vesignieite (with $J \approx 55 K$) and found a plateau at $m = 0.4$ which is a little off from the expected value at $m = \frac{1}{3}$.\cite{Okamoto2011} Another study did observe the $m = \frac{1}{3}$ plateau in another Kagome compound (Cu-titmb). However, the plateau is unstable in Cu-titmb and the nearest neighbor interactions and next nearest neighbor interactions are comparable in this material,\cite{Narumi2004} complicating the physics of the plateaus.

Common approaches used to study frustrated spin systems involve either representing the spin operators in terms of  slave fermions\cite{Baskaran1987, Wen1991} or in terms of slave bosons.\cite{Arovas1988, Sachdev1992} These methods have been used extensively in theoretical and numerical works. Both these approaches work well at the mean-field level but suffer from the limitation that there is no small parameter about which the fluctuations can be treated in a consistent manner. Controlled calculations have been performed by generalizing the $SU(2)$ quantum Heisenberg antiferromagnetic model to an $SU(N)$ or $Sp(N)$ spin model  on the Kagome lattice\cite{Sachdev1992,Wang-2006} but it is not clear if the results obtained in the large-$N$ limit remain valid for the experimentally relevant case of $N = 2$.

Here, we present an alternative approach based on a lattice Chern-Simons theory on a Kagome lattice which implements a two-dimensional Jordan-Wigner transformation that maps hard-core bosons (flipped spins of the antiferromagnet) onto spinless fermions coupled to  the Chern-Simons gauge fields.\cite{Fradkin1989} 
Lattice Chern-Simons theories were defined   for a system on a square lattice,\cite{Fradkin1989,Eliezer1992} and within a consistent lattice Chern-Simons approach, they were used to study  the nearest-neighbor Heisenberg antiferromagnet on the square lattice.\cite{Lopez1994} 

Chern-Simons theories have been very successful in studying and explaining fractional quantum Hall (FQH) type states. These theories yield reliable results in gapped systems and thus may present some new insight into the problem of frustrated quantum systems.  
For these reasons they  have been used for quite some time to study  quantum antiferromagnets on frustrated lattices (triangular, Kagome, 
Shastry-Sutherland, and others) yielding  of intriguing results of possible spin liquid phases.\cite{Yang1993,Misguich2001} In these works these systems were treated only at the level of the average field approximation, and the role of the quantum fluctuations of the Chern-Simons gauge field were ignored. These fluctuations are crucial to the physics of this systems. This is a  well known issue from the analogous theories of the FQH fluids where these fluctuations they play a key role in the physics of the excitations, and in particular their fractional statistics (for a detailed recent discussion of this problem in the FQH fluids see Ref. [\onlinecite{Fradkin2013}].)

However the existing lattice Chern-Simons theory\cite{Fradkin1989} (and its more refined and consistent version by Eliezer and Semenoff\cite{Eliezer1992})   can only be used for systems on a square lattice and, in particular, it cannot be used for frustrated systems on non-bipartite lattices. The Chern-Simons action encodes two key features: 1) the local constraint  requiring the states to be locally gauge invariant (in the form of a Gauss-type law) and 2) a definition of the canonical pairs of fields. The first condition, which for a Chern-Simons theory is a relation between the charge on a site and the gauge flux in an adjacent ``plaquette'', must be obeyed at all sites of the lattice and not just on average. This condition requires that the gauge fluxes on different ``plaquettes'' must commute with each other since otherwise the constraints do not commute with each other (even through while they may still commute locally with the Hamiltonian). This consistency condition sets a restriction on the commutation relations of the gauge fields. Eliezer and Semenoff\cite{Eliezer1992} showed how to impose these conditions consistently for the case of square lattice, at the expense of making the Chern-Simons action less local than would have naively expected.\cite{Fradkin1989} 
It is easy to see that these constraints, even in a non-bipartite lattice,  can only be imposed consistently provided that there is a one-to-one correspondence between sites  and ``plaquettes'' of the lattice. In this paper we will show that this can be done for the (non-bipartite) Kagome lattice. In a separate publication we discuss the generalizations of this construction to more complex lattices. However, this approach does not work for the triangular lattice, for which there are  two different adjacent  triangles (``plaquettes'') for each site of the lattice. Similarly, this procedure does not work for the (bipartite) hexagonal lattice for which two sites are associated with each hexagon.

In this paper we  study the nearest-neighbor $XXZ$ frustrated quantum Heisenberg antiferromagnet on the Kagome lattice using a generalization of the  construction of the Chern-Simons gauge theory of Eliezer and Semenoff to the non-bipartite Kagome lattice which we present here. This construction is one of the main results of this paper. The generalization of this construction for a class of frustrated 2D lattices is presented elsewhere.\cite{Sun2014} In this picture the flipped-spins are represented by hard-core bosons which in turn are described 
as a problem of fermions coupled to a Chern-Simons gauge field on the same lattice. 
Further, the effect of an external magnetic field can also be easily mimicked by adjusting the density of fermions in the equivalent problem. More importantly, this approach will also allow us to go beyond mean-field theory and analyze the effect of fluctuations in such systems. It is here that the effects of the consistent constriction are crucial. 
Here we will focus on the simpler case of the 1/3 magnetization plateau. 

Within our approximations we find that, in the $XY$ limit and for a wide range of the anisotropy parameter $\lambda$, the  ground state of the 1/3 magnetization plateau of the $XXZ$ model on the Kagome lattice is equivalent to a Laughlin fractional quantum Hall state of hard-core bosons with filling fraction $1/2$. This fully gapped state is a topological fluid with a  broken time reversal invariance. This is a state with  spin currents in the ground state and with a fractional Hall spin conductance, has a two-fold ground state degeneracy on the torus, has a single chiral gapless edge state on a disk geometry, and that the excitations of this state are semions. We also found two other plateaus states, one at magnetization $2/3$ (which is equivalent to the $1/3$ plateau) and another one at $5/9$. the plateau at $5/9$ magnetization is also a topological fluid and is equivalent to the first Jain daughter state of the Laughlin FQH state for bosons at filling fraction $2/3$. In this case  the state has two chiral edge states, has a three-fold ground state degeneracy on the torus and the excitations are anyons with statistical angle $\pi/3$. We also showed that, as expected, the spin-spin correlation functions decay exponentially as a function of distance while exhibiting an oscillatory behavior which reflects the breaking of time-reversal symmetry. For large enough anisotropy we find a quantum phase transition to time-reversal invariant states which  extend all the way to the Ising limit, where it has been shown\cite{Cabra2005} that the ground state is governed by a quantum order-by-disorder mechanism and that it equivalent to a complex valence-bond solid. A possible time-reversal invariant  $\mathbb{Z}_2$ spin liquid  has been conjectured to exist at intermediate values of the anisotropy.

The paper is organized as follows. In Section  \ref{sec:Jordan-Wigner_transformation}, the spin-1/2 quantum Heisenberg model on the Kagome lattice  is introduced and the Jordan-Wigner transformation is summarized along with the difficulties related to defining a Chern-Simons term on a lattice. The procedure to obtain a consistent Chern-Simons theory on the Kagome lattice is  described in Subsection \ref{sec:CS_Kagome}. A more detailed discussion on the lattice Chern-Simons theory and when such a procedure works will be presented elsewhere.\cite{Sun2014}  T
In Section \ref{sec:MFT}, we analyze the consequences of this theory at the mean-field level and set up the saddle-point equations for the nearest-neighbor $XXZ$ model after performing the Jordan-Wigner transformation. The mean field theory of the magnetization plateaus is presented in Section \ref{sec:MFT}. 
Subsection~\ref{sec:XY-MFT} deals with the simpler case of the $XY$ regime and the possible magnetization plateaus that can arise are discussed at the mean-field level where this problem reduces  to a problem of interacting fermions hopping on a Kagome lattice in the presence of a background statistical gauge field. This mean field state closely resembles integer quantum Hall type states, typical of the composite fermion approach to the FQH states.
In Subsection \ref{sec:XXZ-MFT} we discuss the  $\frac{1}{3}$, $\frac{2}{3}$ and $\frac{5}{9}$  plateaus, and analyze the full $XXZ$ Heisenberg antiferromagnetic model at these fillings.
 The effect of fluctuations on the $\frac{1}{3}$ and $\frac{5}{9}$ plateau states  are considered in section~\ref{sec:Effects_of_fluctuations}. We discuss the implications of the fluctuations and how they alter the mean-field physics. The states now correspond to a {\em fractional} quantum Hall type state for bosons with fractional (spin) Hall conductivity of $\sigma^s_{xy} = \frac{1}{2}$ and $\frac{2}{3}$ respectively and are thus identified with the Laughlin FQH state for bosons at $\nu=1/2$ and the first Jain state (for bosons) at $\nu=2/3$ . Lastly, we also present  asymptotic calculations of the spin correlations in the magnetization plateaus based on the effective continuum theory ion Section \ref{sec:spin-correlations}. Section \ref{sec:conclusions} is devoted to our conclusions and some open problems. Details of the calculations are relegated to the appendix. We discuss our results and open questions in Section \ref{sec:conclusions}.

\section{Heisenberg model and Jordan-Wigner transformation}
\label{sec:Jordan-Wigner_transformation}

In this section we briefly review the Jordan-Wigner transformation that maps Heisenberg spins to fermions coupled to a Chern-Simons gauge field. This transformation was first discussed in Ref. [\onlinecite{Fradkin1989}]. An important property of the Chern-Simons gauge field is that it imposes a constraint that relates the local density of the fermions to the flux through an adjacent plaquette of the lattice. This flux attachment allows us to identify the spins (which are hard-core bosons) with fermions carrying half a flux quantum. As a result of this constraint, it is crucial to be able to define the Chern-Simons term in a consistent manner on the lattice so that the flux attachment may be performed at each and every site on the lattice. 

The model that we will study is the nearest-neighbor $XXZ$ Heisenberg model on the Kagome lattice in the presence of an external magnetic field $h$
\begin{equation}
	H=J\sum_{<i,j>}\Big[S^x_i S^x_j+S^y_i S^y_j+\lambda S^z_i S^z_j \Big] - h \sum_i S^z_i
	\label{eq:Heisenberg}
\end{equation}
where $J>0$ for antiferromagnetic interactions, $S_i^a$ (with $a=x,y,z$) are the three spin-1/2 operators at lattice site $i$, $<i,j>$ stands for nearest neighboring sites, and $\lambda$ is the magnetic anisotropy parameter along the $z$-direction.

After the Jordan-Wigner transformation, the resultant action with fermions and the Chern-Simons gauge field becomes
\begin{equation}
S=S_{F}(\psi,\psi^{*},A_{\mu})+S_{\rm int}(A_{\mu})+\theta S_{CS}(A_{\mu})
\label{eq:Action_fermions}
\end{equation}
where the fermionic and interacting parts are
	\begin{align}
		S_F(\psi,\psi^*,A_{\mu})  = & \int_t \bigg[ \sum_{x} \psi^*(x,t)\left(iD_{0}+\mu\right)\psi(x,t) \nonumber \\
			& -\frac{J}{2} \sum_{<x,x'>} \left( \psi^*(x,t)e^{iA_j(x,t)}\psi(x',t)+h.c \right) \bigg] 
			\label{eq:Lagrangian-free}\\
		S_{\rm int}(\psi, \psi^*) = & \lambda J \int_t \sum_{<x,x'>}\left(\frac{1}{2} - n(x,t) \right) \left( \frac{1}{2} - n(x',t) \right)
	\label{eq:Lagrangian_fermions}
	\end{align}
where $D_0 = \partial_0 + i A_0$ is the covariant time derivative, $n(x,t) = \psi^*(x,t)\psi(x,t)$ is the fermion density operator (i.e. the site occupancy), and $<x,x'>$ stands for nearest neighboring sites $x$ and $x'$ on the Kagome lattice. Under the  transformation  the $z$ component of the spin operator, $S^z$,  is mapped to the local fermion occupation number 
\begin{equation}
S^z(x,t) = \frac{1}{2} - n(x,t)
\end{equation}
 We can then absorb the external magnetic field term in the Hamiltonian, $h \sum_i S_z^i$, in the definition of the chemical potential $\mu$ in 
 Eq.\eqref{eq:Lagrangian-free}. Hence, the effect of the external magnetic field can easily be mimicked by adjusting the fermionic filling.

In the Jordan-Wigner transformation the parameter $\theta$ is selected so that the statistics of the spins (which are hard-core bosons) are changed in to fermions. This can be done by choosing $\theta = \frac{1}{2\pi (2 k + 1)}$ for any $k \in \mathbb{Z}$. Although the hard-core boson to fermion mapping holds for all integer (positive and negative) values of $k$, we will see below that for two special values, $k=0,-1$ (or, equivalently, $\theta=\pm \frac{1}{2\pi}$), there is a mean field approximation with a fully gapped spectrum. The resulting states for these two choices of $\theta$ are related to each other by time reversal and hence by a reversal of the sign of the magnetization.

In order to complete the Jordan-Wigner transformation, the Chern-Simons term in Eq~\eqref{eq:Action_fermions} needs to be specified. If one naively, extends the continuum version of the Chern-Simons term to a lattice, the flux attachment constraints cannot be imposed consistently as $\left[B(x), B(y) \right] \neq 0$ for any two sites $x$ and $y$ on the lattice.\cite{Fradkin2013} Eliezer and Semenoff \cite{Eliezer1992} developed a form of the  Chern-Simons theory for a square lattice  that can be consistently defined on a square lattice. This lattice Chern-Simons theory was subsequently used to successfully study the (unfrustrated) spin-1/2 quantum Heisenberg antiferromagnet on the square lattice.\cite{Lopez1994}

By generalizing  the procedure outlined by Eliezer and Semenoff, we were able to develop a Chern-Simons theory that can be consistently defined on the non-bipartite Kagome lattice. This will now allow us to use the Jordan-Wigner mapping to study the nearest-neighbor spin-1/2 quantum Heisenberg antiferromagnet on the Kagome lattice. The next section will briefly outline this procedure. A  construction of the lattice Chern-Simons term for more general non-bipartite lattices is presented elsewhere.\cite{Sun2014}

\subsection{Chern-Simons theory on the Kagome lattice}
\label{sec:CS_Kagome}

\begin{figure}[h]
 \includegraphics[width=0.5\textwidth]{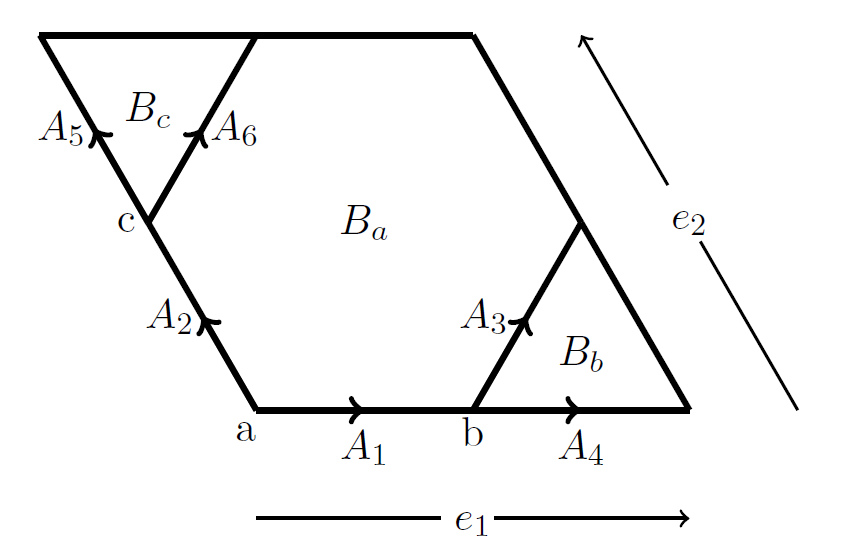}
\caption{Kagome lattice unit cell with three sites attached to the fluxes in their corresponding plaquettes}
\label{fig:Kagome_unit_cell}
\end{figure}

We begin by writing down the below generic form for the lattice Chern-Simons term
\begin{equation}
	\begin{aligned}
 S_{CS} = & S_{CS}^{(1)} + S_{CS}^{(2)} \\
S_{CS}^{(1)} & =  \int dt \sum_{x,y}  A_{0}(x,t)J_{i}(x-y)A_{i}(y,t)  \\
S_{CS}^{(2)} &=  - \frac{1}{2}\int dt \sum_{x,y}A_{i}(x,t)K_{ij}(x-y)\dot{A_{j}}(y,t)
	\end{aligned}
	\label{eq:Generic_CS_term}
\end{equation}
where the $A_0$ fields are defined on the sites of the lattice and the $A_i$ fields are defined on the links of the lattice. See Fig.\ref{fig:Kagome_unit_cell} for our definitions of these gauge fields on the unit cell of the Kagome lattice. Note that in Eq.\eqref{eq:Generic_CS_term} we have omitted the factor of $\theta$.

The first term in Eq.\eqref{eq:Generic_CS_term} is the Gauss law term that imposes the constraint between local density and flux through the plaquettes of the Kagome lattice. The vector kernel $J_i(x-y)$ enforces the condition that relates the charge (i.e. the site occupancy by a fermions) to the gauge flux in the adjacent plaquette. Once a Gauss law has been fixed, the first term in Eq.\eqref{eq:Generic_CS_term} is completely determined. A key feature of the Kagome lattice (shared with the square lattice) is that there is a one-to-one correspondence between sites of the lattice and plaquettes of the same lattice. This condition is not satisfied in other planar lattices, e.g. honeycomb and triangular, which leads to flux attachment prescriptions which are ambiguous and break the symmetries of the lattice. We elaborate more on this issue in Ref.[\onlinecite{Sun2014}].

The structure of the matrix kernel, denoted by $K_{ij}(x-y)$ in the second term in Eq.\eqref{eq:Generic_CS_term}, determines the commutation relations between the different (spatial) gauge fields as follows
\begin{equation}
\left[A_i(x), A_j(x) \right] = - i K_{ij}^{-1}(x-y)
\label{eq:ACommutation}
\end{equation}
It is the structure of the matrix kernel $K_{ij}(x-y)$ in the second term of Eq.\eqref{eq:Generic_CS_term} that will allow us to consistently impose the Gauss law constraints on the lattice. This $K$ matrix also satisfies the condition $K_{ij}(x-y) = -K_{ji}(y-x)$. Since there are six spatial links in the unit cell, this is a $6\times 6$   matrix. {\it The key point is that a lattice Chern-Simons term can be uniquely determined by fixing a Gauss Law, imposing gauge invariance and demanding that the commutation relations between the $A_j$ fields are ``local".} The last condition is primarily included to obtain the simplest form of the lattice Chern-Simons term. Now, we will proceed by establishing these conditions on the Kagome lattice and obtaining the Chern-Simons term. We should note that the matrix $K_{ij}$ of Eq.\eqref{eq:Generic_CS_term} is unrelated to the so-called $K$-matrix that appears in the classification of abelian FQH states.\cite{Wen-1995}
 
\subsection{Gauss' Law} 
\label{sec:gauss}

The first step in writing down the Chern-Simons term is to determine how to impose the Gauss law (flux attachment) on the Kagome lattice. The unit cell of the Kagome lattice has three inequivalent sites (denoted by $a$, $b$ and $c$ in Fig.\ref{fig:Kagome_unit_cell}) and three plaquettes (two triangles, denoted by $b$ and $c$ in Fig.\ref{fig:Kagome_unit_cell}, and one hexagon, denoted by $a$ in Fig.\ref{fig:Kagome_unit_cell}). Thus, there is a natural correspondence between sites and  plaquettes (just as in the case of a square lattice). It turns out that this is the crucial condition that needs to be satisfied in order to write down a Chern-Simons lattice term on a generic lattice.\cite{Sun2014}

On the  Kagome lattice, whose unit cell is shown in Fig.\ref{fig:Kagome_unit_cell}, we  define the flux through each plaquette (defined by the associated sites $a$, $b$ and $c$) to be
\begin{equation}
  \begin{aligned}
      B_a(x) =& A_{1}(x)+A_{3}(x)+A_{5}(x+e_{1}) \\
					 -&A_{4}(x+e_{2})  -A_{6}(x)-A_{2}(x)	\\
      B_b(x) =& A_{4}(x)+A_{2}(x+e_{1})-A_{3}(x)	\\
      B_c(x) =& A_{6}(x)-A_{1}(x+e_{2})-A_{5}(x)
  \end{aligned}
	\label{eq:CS_Gauss_law}
\end{equation}
where $e_1$ and $e_2$ are vectors along the two directions of the lattice as shown in Fig.\ref{fig:Kagome_unit_cell}.
These equations attach the flux at sub-lattice $a$ to that of the hexagon ($B_a$) and the fluxes of sub-lattices $b$ and $c$ to each of the two corresponding triangles ($B_b$ and $B_c$ respectively).

\subsection{Gauge Invariance}
\label{sec:gauge-invariance}

The second step in writing down the Chern-Simons term is to demand that Eq.\eqref{eq:Generic_CS_term} is gauge invariant under the gauge transformations $A_0(x) \mapsto A_0(x) + \partial_0 \Lambda(x)$ for time component of the gauge fields and $A_j(x) \mapsto A_j(x) + \Delta_j \Lambda(x)$  for the space components of the  gauge fields (where $\Delta_j$ is the difference operator on the Kagome lattice). As an example, the field $A_1(x)$ gets transformed as $A_1(x) \mapsto A_1(x) + \Lambda^b(x) - \Lambda^a(x)$ under a gauge transformation where the labels $b$ and $a$ again refer to the sub-lattices and the field $\Lambda(x)$ lives on the sites of the lattice.

The gauge invariance condition is imposed on each of the sub-lattices and is written as
\begin{equation}
		J_{j}^{\alpha}(x-y) + d_{i}^{\alpha}K_{ij}(x-y) = 0
	\label{eq:CS_gauge_invariance}
\end{equation}
where $\alpha = a, b, c$ for each of the sub-lattices. The vectors $J^{\alpha}_j$ (introduced in Eq.\eqref{eq:Generic_CS_term}) and $d^{\alpha}_j$ are most easily written in Fourier space (for the Kagome lattice in Fig.\ref{fig:Kagome_unit_cell}) as
	\begin{align}
		{\bm J}^a(k) & = (1, -1, 1, -e^{-i k_2}, e^{-i k_1}, 1) \nonumber\\
		{\bm J}^b(k) & = (0, e^{-i k_1}, -1, 1, 0, 0) \nonumber\\
		{\bm J}^c(k) & = (-e^{-i k_2}, 0, 0, 0, -1, 1) 
		\label{eq:Gauss_law_J1}
	\end{align}
and
	\begin{align}
		{\bm d}^a(k) & = (-1, -1, 0, e^{i k_1}, e^{i k_2}, 0) \nonumber\\
		{\bm d}^b(k) & = (1, 0, -1, -1, 0, e^{i k_2}) \nonumber\\
		{\bm d}^c(k) & = (0, 1, e^{i k_1}, 0, -1, -1) 
		\label{eq:Gauss_law_J2}
	\end{align}

\subsection{Local Commutation Relations}
\label{sec:local-CCR}

\begin{figure}
                \includegraphics[width=0.4\textwidth]{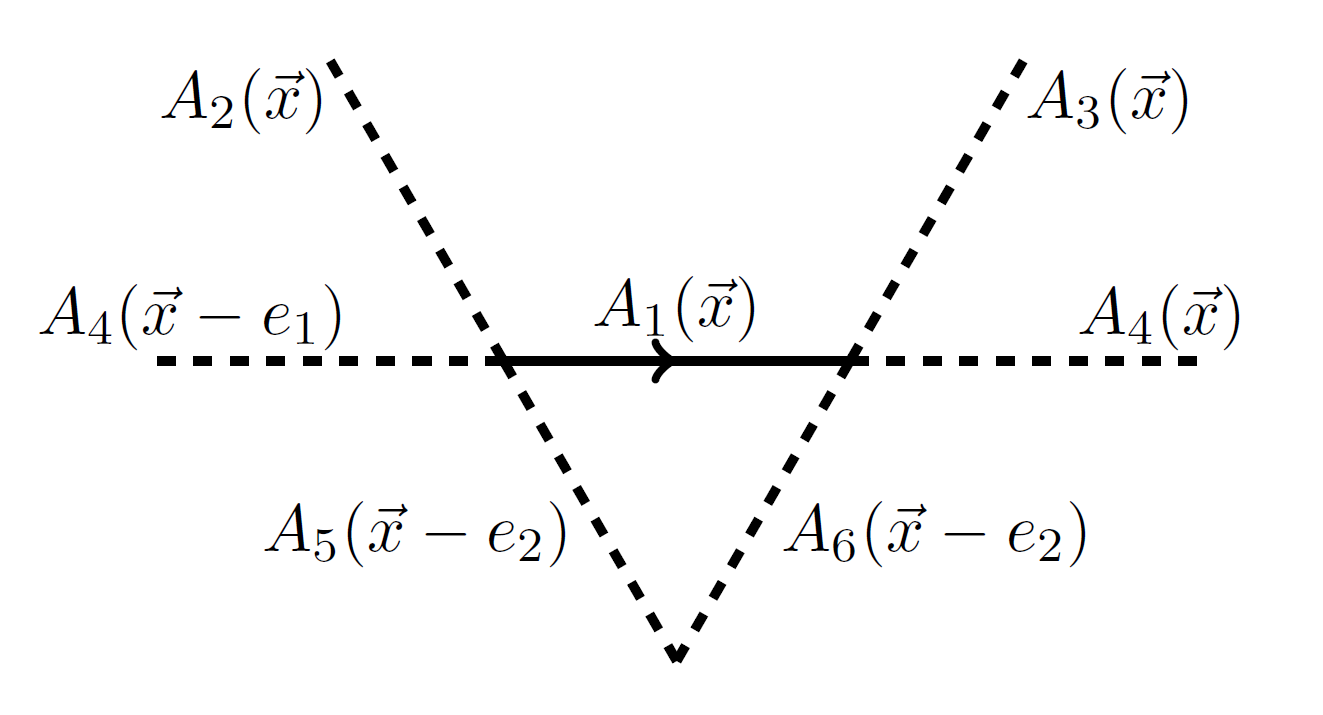}
        \caption{The field $A_1$ has non-trivial commutation relations with only these six links. This is the ``locality" condition that is imposed.}
		\label{fig:A1_commutation}
\end{figure}

So far we have established the Gauss Law and imposed gauge invariance. The final step is to look for a form of the matrix $K$ that is ``local" i.e. only links (spatial gauge fields $A_j$) that touch one another have non-trivial commutation relations as given by Eq.\eqref{eq:ACommutation}. This is most easily seen in Fig.\ref{fig:A1_commutation} where the link $A_1$ is shown as an example. The link $A_1(x)$ has non-trivial commutation relations with just the six links that it touches and it commutes with all the other links on the lattice.  

Using the above conditions one can obtain the below unique form of the matrix $K$  
\begin{widetext}
\begin{equation}
K = 
\frac{1}{2} \left(
\begin{array}{cccccc}
 0 & -1 & 1 & -S_2 & S_1+S_2^{-1} & -1+S_2^{-1} \\
 1 & 0 & 1-S_1^{-1} & -S_2-S_1^{-1} & S_1 & -1 \\
 -1 & S_1-1 & 0 & 1-S_2 & S_1 & -1 \\
 S_2^{-1} & S_1+S_2^{-1} & S_2^{-1}-1 & 0 & S_1 S_2^{-1} & S_2^{-1} \\
 -S_2-S_1^{-1} & -S_1^{-1} & -S_1^{-1} & -S_2S_1^{-1} & 0 & 1-S_1^{-1} \\
 1-S_2 & 1 & 1 & -S_2 & S_1-1 & 0 \\
\end{array}
\right)
	\label{eq:K_matrix_Kagome}
\end{equation}
\end{widetext}
where $S_j$ are {\it lattice} shift operators along the two different directions ($e_1$ and $e_2$) on the lattice i.e. $S_j f(x) = f(x+e_j)$.  Also since $\textrm{Det} K = 1$, the matrix $K$ is invertible.

The above form of the matrix $K$ in Eq.\eqref{eq:K_matrix_Kagome} ensures that the fluxes  commute with each other for any pair of sites $x$ and $y$ on the Kagome lattice (i.e. $[B(x), B(y)] = 0$ for any $x$ and $y$). This will allow us to impose the Gauss Law constraint consistently on each and every site of the Kagome lattice. More precisely, any two Wilson loops on the lattice commute, and Wilson lines on the lattice obey a signed intersection rule analogous to what happens in the continuum case of a Chern-Simons theory. This completes the mapping 
of Eq.\eqref{eq:Action_fermions} for the nearest-neighbor $XXZ$ Heisenberg antiferromagnet on the Kagome lattice into a system of fermions coupled to Chern-Simons gauge fields.
A more detailed discussion of the construction of the Chern-Simons theory on other non-bipartite lattices will be presented in another publication.\cite{Sun2014}  Here, we continue by looking at the consequences of this theory on the Kagome lattice.

\subsection{Jordan-Wigner transformation}
\label{sec:JW}

In the preceding subsections we showed that a system of hard-core bosons on a kagome lattice, representing the flipped spins of the spin-1/2 anisotropic Heisenberg antiferromagnet on the kagome lattice (relative to the uniformly magnetized reference state), is equivalent to a system of fermions on the same kagome lattice minimally coupled to a Chern-Simons gauge field (defined on the links of the kagome lattice). The same result was obtained earlier on for the case of the square lattice.\cite{Fradkin1989,Lopez1994,Eliezer1992,Eliezer1992a} 

A consequence of this mapping (actually, an identity) is the operator identification
\begin{equation}
S_z(\bm x)=\frac{1}{2}-\theta B(\bm x)
\label{eq:Sz}
\end{equation}
which follows from the local (Gauss law) constraint of the Chern-Simons theory, discussed in subsection \ref{sec:gauss},  which relates the {\em local} fermion occupation number of lattice site $\bm x$ to the flux $B(\bm x)$ on the adjoining plaquette.

A second consequence is the following formal identification of the spin raising and lowering operators $S^\pm({\bm x})$ in terms of fermion operators coupled to Wilson lines of the gauge field
\begin{equation}
S^+(\bm x)=\psi^\dagger(\bm x) \; e^{i \sum_{\gamma(\bm x)} A}, \quad S^-(\bm x)= e^{-i \sum_{\gamma(\bm x)} A}\; \psi(\bm x) 
\label{eq:Spm}
\end{equation}
where the sum in the exponent has to be interpreted as the oriented sum of gauge fields defined on the links of the  lattice on an (arbitrary) open path $\gamma(\bm x)$ ending at the site $\bm x$. The operators defined on the right hand sides of Eq.\eqref{eq:Spm} are manifestly gauge invariant and square to zero, and using the commutation relations of the Chern-Simons gauge fields, they are found to obey bosonic commutation relations (provided $\theta=\frac{1}{2\pi}$). For a system with the geometry of a disk, the exponential factors can equivalently be rewritten in the form 
 \begin{equation}
S^+(\bm x)=\psi^\dagger(\bm x) \; e^{i \Phi(\bm x)}, \quad S^-(\bm x)= e^{-i \Phi(\bm x)}\; \psi(\bm x) 
\label{eq:Spm2}
\end{equation}
where the operator $\exp(i \Phi(\bm x))$ is a {\em disorder operator} that creates a fluxoid of strength $1/\theta$ at the plaquette adjoining site $\bm x$ of the Kagome lattice. This is the generalized Jordan-Wigner transformation of Ref.[\onlinecite{Fradkin1989}].

\section{Mean-field theory of the magnetization plateaus}
\label{sec:MFT}

In this section we discuss the physics of the magnetization plateaus of the spin-1/2 $XXZ$ quantum Heisenberg antiferromagnet on the Kagome lattice  at the mean-field level. As already mentioned above the Chern-Simons term imposes the flux attachment condition $n(x,t) = \theta B(x,t)$. This allows us to re-write the interaction term of the action for the fermions, {\em c.f.} Eq.\eqref{eq:Lagrangian_fermions}, (originally the $S_z S_z$ term of the Heisenberg Hamiltonian) purely in terms of the Chern-Simons gauge field as follows
\begin{equation}
	S_{\rm int}(A_{\mu}) = \int dt \lambda J \sum_{<x,y>}\left(\frac{1}{2} - \theta B(x,t) \right) \left( \frac{1}{2} - \theta B(y,t) \right)
	\label{eq:Interaction_A}
\end{equation}
As a result of this substitution, the action is now quadratic in fermionic fields and the fermionic degrees of freedom can be integrated out to yield the effective action just in terms of gauge fields. The effective action has the form
\begin{equation}
S_{\rm eff}(A_{\mu})=-i \textrm{tr}\ln[iD_{0}+\mu-h(A)]+S_{\rm int}(A_{\mu})+\theta S_{CS}(A_{\mu})
	\label{eq:Action_gauge_fields}
\end{equation}
where the hopping Hamiltonian $h(A)$ is (in matrix notation)
\begin{equation}
	h(A)=\frac{J}{2}\sum_{<x;x'> }\left[e^{iA_{j}(x,t)} \ket{x,t} \bra{x',t}+h.c\right] 
	\label{eq:Hamiltonian_matrix}
\end{equation}
and the label $<x,x'> $ refers to the nearest neighbors sites $x$ and $x'$ on the Kagome lattice.

\subsection{Saddle-point equations}
\label{sec:Mean-Field_equations}

The saddle-point equations are obtained by extremizing the action in Eq.\eqref{eq:Action_gauge_fields} w.r.t the gauge fields 
\begin{equation}
\frac{\delta S_{\rm eff}(A)}{\delta A_{\mu}}\Bigg|_{A_{\mu}=\bar{A}_{\mu}}=0
\end{equation}
Before writing down the expressions for the saddle point equations, we first focus on the fermionic part of the action $S_F$. Its derivative w.r.t the temporal $A_0$ component of the  gauge field gives
\begin{equation}
\langle n(x,t) \rangle=\left\langle -\frac{\delta S_{F}}{\delta A_{0}(x,t)}\right\rangle =-iS(x,t;x,t)
\end{equation}
Similarly, for the spatial $A_k$ component one gets
\begin{equation}
	\begin{aligned}
\langle j_{k}(x,t) \rangle= &  \left\langle -\frac{\delta S_{F}}{\delta A_{k}(x,t)}\right\rangle \\ 
				 			=   \frac{J}{2} & \bigg[  S(x+e_{k},t;x,t)e^{i\bar{A}_{k}(x,t)}  \\
								& -S(x,t;x+e_{j},t)e^{-i\bar{A}_{k}(x,t)} \bigg]
	\end{aligned}
\end{equation}
Here, $ j_{k}(x,t)$ is the (gauge-invariant) fermionic current. $S(x,t;x',t')$ is the propagator  for the fermions in an average background field $\bar{A}_{\mu}(x,t)$, and it is the solution of the lattice differential equation 
\begin{equation}
\left(i\bar{D}_{0}+\mu-h(\bar{A})\right)S(x,t;x',t')=\delta_{x,x'}\delta(t-t')
\label{eq:Green_function_MF}
\end{equation}
where $h(\bar{A})$ is given in Eq.\eqref{eq:Hamiltonian_matrix}.

Hence, the saddle point equation w.r.t $A_{0}(x)$ field yields the expectation value of the local fermion density
\begin{equation}
	\langle n(x) \rangle=\theta \langle B(x) \rangle
	\label{eq:MF_density}
\end{equation}
which  amounts to imposing the flux attachment constraint on average.

Similarly for the field $A_{k}(x)$, one gets the mean-field equation  for the expectation value of the local fermion current
\begin{equation}
\left\langle j_{k}(x,t)\right\rangle =\theta\left\langle \frac{\delta S_{CS}}{\delta A_{k}(x,t)}\right\rangle +\left\langle \frac{\delta S_{\rm int}}{\delta A_{k}(x,t)}\right\rangle 
\label{eq:<j>}
\end{equation}
The expectation values in Eq.\eqref{eq:<j>} are explicitly given by
\begin{equation}
	\begin{aligned}
	\left\langle \frac{\delta S_{CS}^{(1)}}{\delta A_{k}(x,t)} \right\rangle =& \bar{d}^{k\alpha} \bar{A}_{0\alpha}(x) \\
\left\langle \frac{\delta S_{CS}^{(2)}}{\delta A_{k}(x,t)}\right\rangle =&\frac{1}{2}(K_{ki}-K_{ik})\partial_{0}\bar{A}_{i}(x) = 
  		K_{ki}\partial_{0}\bar{A}_{i}(x)
	\end{aligned}
\end{equation}
where $\alpha$ is the sub-lattice index and
\begin{equation}
\bar{d}^{k\alpha} = 
\left(
\begin{array}{ccc}
1 & 0 & -S_2^{-1} \\
-1 & S_1^{-1} & 0	\\
1 & -1 & 0 \\
-S_2^{-1} & 1 & 0 \\
S_1^{-1} & 0 & -1 \\
-1 & 0  & 1
\end{array}
\right)
\end{equation}
where $S_1$ and $S_2$ are again the shift operators as defined earlier in Section~\ref{sec:CS_Kagome}. 

The full form of the saddle point-equation for the $A_k$ fields is quite cumbersome and will not be written down explicitly here. Instead, as we are looking for time-independent/static and uniform solutions, we take the fluxes on any particular sub-lattice to be the same (i.e. $\bar{B}^{\alpha}(x)=\bar{B}^{\alpha}(y)$ for any $x$ and $y$), and the resulting simplified mean-field expression for the mean-field currents is
\begin{equation}
\begin{aligned}
	\langle j_{k}(x) \rangle= & \theta \bar{d}^{k\alpha}\bar{A}_{0\alpha}(x) \\
		&	 -2J\lambda\theta^{2} (-1)^{k} \left[ \bar{B}^{a} - f_{k} \bar{B}^{c} - (1-f_{k}) \bar{B}^{b} \right]
\end{aligned}
\label{eq:MF_current}
\end{equation}
with $f_k = 1$ when $k = 1, 5, 6$ and $f_k = 0$ when $k = 2, 3, 4$. 

The expressions for the mean-field state were derived by assuming translation invariance and allowing for  each sublattice to be inequivalent from the others. If one is looking for other types of states (that break translational symmetry for example), then the mean-field expressions would have to be modified accordingly.


\subsection{$XY$ model}
\label{sec:XY-MFT}

Let us now analyze the $XY$ Heisenberg antiferromagnet $(\lambda = 0)$ and its magnetization plateaus at the mean-field level. Setting $\lambda = 0$ gets rid of the $S^z S^z$ components and makes the fermions non-interacting. However, the fermions are still coupled to the Chern-Simons gauge field. Further, the mean-field equation Eq.\eqref{eq:MF_current} is satisfied by $\bar{A}_{0\alpha} = 0$ in the absence of any currents. This implies that we are just left with the flux attachment condition in Eq.\eqref{eq:MF_density}.

At the mean-field level, we look for uniform flux states i.e. $\bar{B}_a = \bar{B}_b = \bar{B}_c = \phi=2\pi  \frac{p}{q}$ with $p, q \in \mathbb{Z}$. This makes the total flux through the unit cell (which has three plaquettes) $B_{u.c}=3\phi =2\pi  \frac{3p}{q}$. By imposing the Chern-Simons constraint on average, we deduce that, for uniform states and taking into account that  $\theta=\frac{1}{2\pi}$,  the average site occupancy (density) of each sublattice of the unit cell is $\langle n \rangle=\frac{p}{q}$.

Such a state can be realized with the below choice of gauge fields 
\begin{equation}
	\begin{matrix}
		A_{1}(\vec{x})=0  & & A_{2}(\vec{x})=\phi  & &A_{3}(\vec{x})=0 \\
		A_{4}(\vec{x})=0  & &  A_{5}(\vec{x})=-\phi+3\phi x_{1} && A_{6}(\vec{x})=3\phi x_{1}
		\end{matrix}
\end{equation}
with $\vec{x} = (x_1, x_2)$ where $x_1$ and $x_2$ are the co-ordinates along the $e_1$ and $e_2$ directions respectively in Fig.\ref{fig:Kagome_unit_cell}.

\subsubsection{Hofstadter spectrum}
\label{sec:XY-spectrum}

The $XY$ model has now been reduced to a problem of non-interacting fermions hopping in a lattice in the presence of a (statistical) magnetic field. This is very similar to the problem of the integer quantum Hall (IQH) effect where the one-particle states possess non-trivial Chern numbers. For a square lattice one can obtain these Chern numbers by solving  the resulting Harper equation either numerically or by performing a perturbation theory in the hopping parameters.\cite{Thouless1982} The final structure is most easily seen in the Hofstadter spectrum as was pointed out by G. Misguich et. al. \cite{Misguich2001} in their studies on the triangular and Shastry-Sutherland lattices using a similar analysis. By extrapolating the Chern numbers from the case of the square lattice, one can obtain the Chern numbers for the case of the Kagome $XY$ Heisenberg model. 

\begin{figure}[hbt]
\includegraphics[width=0.5\textwidth]{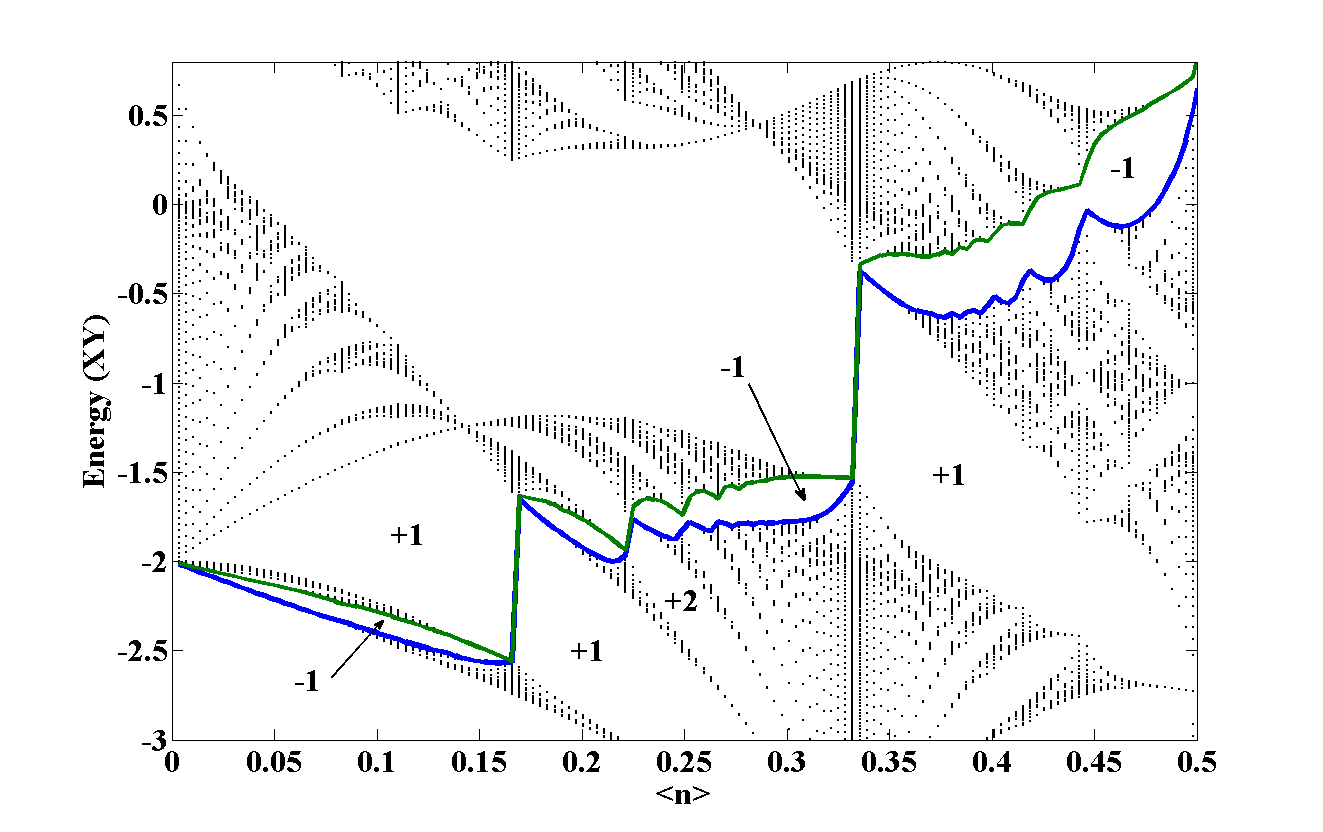}
\caption{Hofstadter spectrum  for the $XY$ model as a function of sublattice density $\langle n \rangle$. The numbers shown are the Chern numbers of the respective filled bands. The magnetization plateaus at $\frac{1}{3}, \frac{5}{9}, \frac{2}{3}$ correspond to the three vertical jumps shown in this figure, respectively at densities $\langle n \rangle=\frac{1}{3}, \frac{2}{9},\frac{1}{6}$.}
\label{fig:Hofstadter}
\end{figure}

The results are shown in Fig.\ref{fig:Hofstadter} where the $x$-axis is the average filling/density $\langle n \rangle$ on each sub-lattice and the $y$-axis are the single-particle energies of the associated free fermion model of the $XY$ model. The bottom solid line indicates the Fermi level for the occupied bands. The top solid line is the next excited energy single-particle state available.

The numbers shown in Fig.\ref{fig:Hofstadter} are the Chern numbers of a state with all the below bands completely filled up to that number. In Section \ref{sec:Full_continuum_action}  we will see  that  we have a quantum Hall type incompressible state if the total Chern number of the occupied bands satisfies $C\neq -1$.
The discontinuous jumps in the Fermi level in the figure indicate the fillings at which the Chern number $C \neq -1$. These situations are expected to correspond to magnetization plateaus.\cite{Misguich2001}
These jumps occur at site fillings of $\langle n \rangle = \frac{1}{3}, \frac{1}{6}, \frac{2}{9}$, which correspond, respectively, to $2\pi$ flux through each unit cell, $\pi$ flux through each unit cell (and hence  $2\pi$ flux for two unit cells), and $\frac{4\pi}{3}$ flux per unit cell (or $4\pi$ flux in three unit cells). Since
\begin{equation}
	\langle n \rangle = \frac{1}{2} - \langle S_z \rangle = \frac{1}{2} - M
\end{equation}
we could find possible magnetization plateaus corresponding to $m = \left| \frac{M}{M_{sat}} \right|= \frac{1}{3}, \frac{2}{3}$ and $\frac{5}{9}$, with $M_{sat} = \frac{1}{2}$. Hence, at the mean-field level all these magnetization  states have integer Chern numbers and behave like integer quantum Hall (IQH) states, much in the same way as with the behavior of composite fermions in the theory of the fractional quantum Hall effect.\cite{Jain1989,Jain1992,Lopez1991} In what follows we will focus primarily on the simplest case of the $1/3$ magnetization plateau and comment briefly on the other cases.

\subsection{$XXZ$ model}
\label{sec:XXZ-MFT}

Following our discussion in the previous section, we now  extend the results to the case of the $XXZ$ model by introducing the $\lambda$ term $(S^z S^z)$ in the Hamiltonian. We begin first by performing the mean-field analysis at the fillings associated with the magnetization plateaus.

\subsubsection{Mean-field analysis}
\label{sec:MFT-XXZ}

At the mean-field level each unit cell satisfies the condition $\langle n \rangle_{u.c} = \langle n_a \rangle + \langle n_b \rangle + \langle n_c \rangle = 1$. This density condition translates to the flux condition $\langle B_a \rangle + \langle B_b \rangle + \langle B_c \rangle = 2\pi$ as a result of the flux attachment constraint in Eq.\eqref{eq:MF_density}. (Fluxes on the lattice are defined modulo $2\pi$.) The $B$ fields are gauge invariant quantities and the above ansatz can be satisfied by the below choice of gauge fields  
\begin{equation}
	\begin{matrix}
		\bar{A}_1=-c_2		&	&  \bar{A}_2=c_1		&	&	\bar{A}_3=-c_1	\\
		\bar{A}_4=c_1		&	&	\bar{A}_5=-c_2		& 	&	\bar{A}_6=c_2
	\end{matrix}
	\label{eq:MF_spatial_gauges}
\end{equation}
where $c_1$ and $c_2$ are some constant parameters that will be determined below and the definitions of the links on the Kagome lattice in Fig.\ref{fig:Kagome_unit_cell} are used. This makes 
\begin{equation}
	\begin{matrix}
		\bar{B}_a= 2\pi-3c_1-3c_2		&	&	\bar{B}_b=3c_1		&	&	\bar{B}_c=3 c_2
	\end{matrix}
	\label{eq:MF_fluxes}
\end{equation}

Further, assuming that these ground states states have no currents i.e. $\langle j_k(x,t) \rangle = 0$, the second saddle point equation Eq.\eqref{eq:MF_current} can be satisfied by the below choice of temporal gauge fields on each of the sublattices
\begin{equation}
	\begin{aligned}
		\bar{A}^a_0 & =-2J \lambda  \theta  \left(2\pi -3c_1-3c_2\right) \\
		\bar{A}^b_0 & =2 J \lambda  \theta  3c_1 \\
		\bar{A}^c_0 & =2 J \lambda  \theta  3c_2
	\end{aligned}	
	\label{eq:MF_temporal_gauges}
\end{equation}

\subsubsection{Self-consistent solutions}
\label{sec:Self-consistent_solutions}

The parameters $c_1$ and $c_2$ can be computed (numerically) by demanding that the saddle-point equation Eq.\eqref{eq:MF_density} is satisfied on each sub-lattice for a given value of the magnetic anisotropy parameter $\lambda$, i.e. 
\begin{equation}
		\langle n_{\alpha}(x,t) \rangle = -i S_{\alpha, \alpha}(x,t;x,t)
	\label{eq:Self_consistency}
\end{equation}
where $\alpha$ is the sub-lattice index.
The expression for the propagator in momentum space is given by
\begin{equation}
	S_{\alpha \beta}(x,t; x',t') = \int_{\omega, \vec{k} \in B.Z.} e^{i \omega(t-t')-i k(x-x')} S_{\alpha \beta}(\omega, \vec{k})
\end{equation}
where
\begin{widetext}
	\begin{equation}
		S_{\alpha \beta}^{-1}(\omega, \vec{k})=\left(
\begin{array}{ccc}
 \omega -\bar{A}_{0 a} & -\frac{J}{2} \left(e^{-i\bar{A}_4-i k_1}+e^{i \bar{A}_1}\right) & -\frac{J}{2} \left(e^{-i \bar{A}_5-i k_2}+e^{i \bar{A}_2}\right) \\
 -\frac{J}{2} \left(e^{i \bar{A}_4+i k_1}+e^{-i \bar{A}_1}\right) & \omega -\bar{A}_{0 b} & -\frac{J}{2} \left(e^{i \bar{A}_3+i k_1}+e^{-i \bar{A}_6-i k_2}\right) \\
 -\frac{J}{2} \left(e^{i \bar{A}_5+i k_2}+e^{-i \bar{A}_2}\right) & -\frac{J}{2} \left(e^{i \bar{A}_6+i k_2}+e^{-i \bar{A}_3-i k_1}\right) & \omega -\bar{A}_{0 c} \\
\end{array}
\right)
	\label{eq:MF_propagator}
	\end{equation}
\end{widetext}
The values of $c_1$ and $c_2$ that satisfy Eq.\eqref{eq:Self_consistency} are listed in table \ref{MF_consistency} for a few different values of $\lambda$. For the $XY$ model, $\lambda = 0$, the densities on all the sites are the same (and equal to 1/3). As $\lambda$ is increased (and the interactions are turned on),  the density $n_a$ steadily increases while the density $n_b=n_c$ decreases. For very large anisotropy $\lambda$, $n_a \approx 1$ while $n_b = n_b \approx 0$ in this model. Intuitively, in the Ising limit ($\lambda \to \infty$), this corresponds to the spins either pointing strictly up or down as expected. At one-third filling, this translates to two up spins and one down spin on average. The fluxes on each of the plaquettes would then either be $0$ or $2\pi$ which are equivalent on the lattice. Hence, in the Ising limit, this maps to a problem of fermions hopping on the Kagome lattice with no flux at the mean-field level. 

\subsubsection{Chern Numbers of the Hofstadter States}

\begin{figure}[hbt]
  \begin{center}
  \subfigure[$\lambda=0$]{  \includegraphics[width=0.45\textwidth]{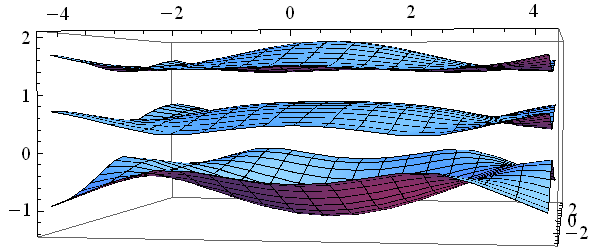}}
        	\\
\subfigure[$\lambda=0.6$]{  \includegraphics[width=0.45\textwidth]{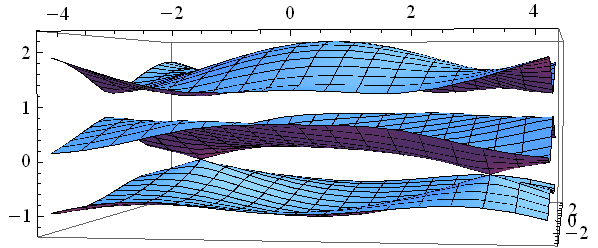}}
\end{center}
		\caption{The Mean-Field spectrum at $\frac{1}{3}$ filling for $\lambda=0$ and $\lambda=0.6$. At $\frac{1}{3}$ filling only the bottom band is filled. The spectrum is gapped for all values of $\lambda$ except for $\lambda \approx 0.6$ where the bottom band touches the middle band and the Chern numbers of the corresponding bands switch.}	
		\label{fig:MF_spec_int}
\end{figure}

The energy spectrum of the mean-field phases has three bands as shown in Fig.\ref{fig:MF_spec_int} (for two values of anisotropy parameter $\lambda$). At $\frac{1}{3}$ filling only the bottom band is filled. This state is gapped for all values of $\lambda$ except at $\lambda \approx 0.6$ when the bottom two bands cross and the low-energy fermionic states become gapless Dirac fermions. 

\begin{table}
	\begin{tabular}{|c|c|c|c|}
		\hline
		 $\lambda$  & $c_1=c_2$  &$n_a$ & $n_b=n_c$ \\
		\hline 
		0 & 0.698062 & 0.3333 & 0.3333 \\
		0.25 & 0.600673 & 0.4264 & 0.2868 \\
		0.5 & 0.450504 & 0.5698 & 0.2151 \\ 		
		0.6 & 0.361283 & 0.655 & 0.1725 \\
		0.75 & 0.235619 & 0.775 & 0.1125 \\
		1 & 0.143257 & 0.8632 & 0.0684 \\
		\hline
	\end{tabular}
	\caption{Sublattice occupation numbers $n_a$, $n_b$ and $n_c$, and values of the parameters $c_1=c_2$ for different values of the magnetic anisotropy $\lambda$, where $\lambda=0$ is the $XY$ model and $\lambda \to \infty$ is the Ising model.}
	\label{MF_consistency}
\end{table}

The Chern number of the bands is given by
\begin{equation}
  C = \frac{1}{2\pi} \int_{BZ} d^{2}k F_{12}(k)
\end{equation}
where $F_{ij} = \partial_{i}a_{j} - \partial_{j}a_{i}$ and $a_{i} = -i \bra{\psi}\partial_{k_{i}} \ket{\psi}$ is the Berry flux. Here $\ket{\psi}$ is the normalized eigenvector of the corresponding single-particle energy band and the integral is over the first Brillouin zone of the Kagome lattice.
 The Chern numbers of the three bands are shown in   table \ref{Chern_no}.
\begin{table}[hbt]
\begin{tabular}{|c|c|c|c|}
	\hline
	$\lambda$ & $C_{bottom}$ & $C_{middle}$ & $C_{top}$ \\
	\hline
	$\lesssim$ 0.6 & +1 & 0 & -1 \\
	$\gtrsim$ 0.6 & 0 & +1 & -1 \\
	\hline
\end{tabular}
\caption{Chern numbers for the bottom, middle and top bands for $\lambda <0.6$ and $\lambda > 0.6$ at $1/3$ filling.}
\label{Chern_no}
\end{table}
At $\lambda = 0$, the Chern number in table \ref{Chern_no} matches the result for the Chern number obtained from the Hofstadter spectrum shown in Fig.
 \ref{fig:Hofstadter} at $\frac{1}{3}$ filling. Hence, once again at the mean-field level and for $\lambda \lesssim 0.6$, the phase is gapped and looks like an integer quantum Hall  state with Chern number $C_{bottom} = +1$. The point $\lambda \sim 0.6$ marks a transition point between two different phases, an IQH state and an insulating state, again at the mean-field level.

As we had noted in our introductory section on the Jordan-Wigner transformation, the above analysis is valid for $\theta = \frac{1}{2\pi}$. But we could just as easily have chosen $\theta = -\frac{1}{2\pi}$. For this choice of $\theta$ the fluxes on each of the plaquettes pick up a negative sign and the sublattice magnetizations have now the opposite sign. This would yield a mean-field state that is related to the above mean-field state by time-reversal symmetry. Also all the signs of the Chern numbers in Table \ref{Chern_no} would be flipped. We would still get an integer quantum Hall state but with opposite Chern number $C = -1$.

\section{Continuum Effective Action for the Magnetization Plateaus}
\label{sec:eff-action-1/3}

We  now turn  to  study the effects of quantum fluctuations for the magnetization plateaus. From now on we will now focus on the case of $\langle n \rangle = \frac{1}{3}$ filling. In this case the magnetic unit cell is the same as the regular unit cell of the Kagome lattice making it the simplest case to study analytically. Although the other plateaus require  larger magnetic unit cells and make the computation analytically unfeasible, the leading part of the long-distance effective action, i.e. the topological piece, can be computed for all three plateaus.

After performing the Jordan-Wigner transformation one obtains the action given by Eq.\eqref{eq:Action_fermions} and Eq.\eqref{eq:Lagrangian_fermions} which is reproduced here
\begin{align}
S = \int dt \sum _x & \big\{ \psi ^*(x)\left[i D_0+\mu \right]\psi (x) \nonumber \\
		  - \sum _{<x,x'>}  \frac{J}{2} & \psi ^*(x)e^{i A_j(x)}\psi \left(x+e_j\right)+c.c\big\}+S_{\rm int}+\theta
 S_{CS}
\end{align}
with $D_0=\partial _0+i A_0$ and again $<x,x'>$ refers to pairs of nearest neighbor sites of the Kagome lattice. Once again at the mean-field level, the state is described as given by Eq. \eqref{eq:MF_spatial_gauges}, Eq. \eqref{eq:MF_temporal_gauges} and in Table \ref{MF_consistency}. The corresponding mean-field phases were discussed in the previous section \ref{sec:Self-consistent_solutions}. Now, we analyze the effect of fluctuations about this mean-field state.

\subsection{Fluctuations and the $1/3$ plateau}
\label{sec:Effects_of_fluctuations}

In this subsection, we perform an expansion around the mean-field state of the $XY$ model and of the $XXZ$ model. As was outlined earlier in Section \ref{sec:Mean-Field_equations}, the action is quadratic in fermions and can be integrated out to give Eq.\eqref{eq:Action_gauge_fields}. This allows us to perform an expansion around the mean-field state in powers of the fluctuations of the gauge fields by expressing $A_{\mu } = \bar{A}_{\mu }+\delta A_{\mu }$. Here $\bar{A}_{\mu}$ correspond to the mean-field values of the gauge fields in Eq.\eqref{eq:MF_spatial_gauges} and Eq.\eqref{eq:MF_temporal_gauges}. The final action in the terms of the fluctuating components will have the form 
\begin{equation}
	S_{\rm eff} =\frac{1}{2} \int d^3x d^3y \delta A_{\mu}(x)\Pi^{\mu \nu}(x,y)\delta A_{\nu}(y) + \theta S_{CS}+ S_{\rm int}
	\label{eq:Action_polarization}
\end{equation}
where $\Pi^{\mu\nu}$ is the polarization tensor and $\delta A_{\mu}$ are the fluctuations of the gauge fields. However, unlike the conventional polarization tensor in $2+1$ dimensions where $\mu = 0, 1, 2$, now the indices $\mu$ can take a total of nine possible values corresponding to the three temporal fluctuation components and the six spatial fluctuation components on the unit cell of the Kagome lattice. 

One way to reduce these additional degrees of freedom is to express the fluctuating components in terms of slow and fast components as follows
\begin{equation}
	\begin{matrix}
\delta A_{0a}=\delta A_{0s}+\delta A^{f1}_0+\delta  A^{f2}_0 \\
\delta A_{0b}=\delta A_{0s}-\delta A^{f1}_0+\delta  A^{f2}_0 \\
\delta A_{0c}=\delta A_{0s}+\delta A^{f1}_0-\delta A^{f2}_0 
	\end{matrix}
	\label{eq:Temporal_fluctuations}
\end{equation} 
where the labels $a, b, c$ refer to the sub-lattice indices. The subscript label $0s$ refers to the slow component of the temporal fluctuations and the superscript labels $f1$ and $f2$ refer to the fast components of the temporal fluctuations. (In the absence of the fast components, this just amounts to replacing the various fluctuations on the sub-lattices with a slowly varying fluctuating component.) This construction allows us to treat the slow fluctuations as the more relevant fields.
We will also show below that it is possible to integrate out all the fast fluctuating components and eventually obtain an effective action that only depends on the slow fields. 

Similarly for the spatial gauge fields
\begin{equation}
	\begin{matrix}
	\delta A_1=\delta A_{1s}+\delta A^f_1  & & 
	\delta A_4=\delta A_{1s}-\delta A^f_1 \\
	\delta A_2=\delta A_{2s}-\delta A^f_2 & & 
	\delta A_5=\delta A_{2s}+\delta A^f_2 \\
	\delta A_3=\delta A_{3s}-\delta A^f_3 & & 
	\delta A_6=\delta A_{3s}+\delta A^f_3	
	\end{matrix}
	\label{eq:Spatial_fluctuations}
\end{equation} 
where the subscript labels $j s$ refer to the slow components and the superscript labels $f$ refer to the fast components again. To simplify the notation, the $\delta$ label for fluctuations and the label $s$ for the slow components will be dropped from now on. 

The polarization tensor $\Pi^{\mu\nu}$ in Eq.\eqref{eq:Action_polarization} is calculated by computing the one loop correction about the mean-field state using the mean-field fermion Green function in  Eq.\eqref{eq:MF_propagator}. This computation will be performed numerically due to the complexity of the energy bands in the mean-field phase. The final action in Eq.\eqref{eq:Action_polarization} must also be  invariant under local gauge transformations, $A_{\mu }\to  A_{\mu }+\partial_{\mu }\Lambda$. This imposes the transversality condition on the polarization tensor,  $\partial ^x_{\mu }\Pi _{\mu \nu }(x,y)=0$ or equivalently, $P^{\mu }\Pi _{\mu \nu }(P)=0$ in Fourier space (under $A_{\mu }(P)\to  A_{\mu }(P)-i P_{\mu }\lambda (P)$) on the polarization tensor. The transversality condition can be used to simplify the computation to some extent. More explicit details of this calculation are shown in Appendix \ref{app:Continuum_Action}

\subsection{Full Continuum Action}
\label{sec:Full_continuum_action}

Expanding the above action about the mean-field state up to second order in fluctuations gives the below terms
\begin{equation}
S_{\rm final} = S_{00} + S_{0j} + S_{ij} + \theta S_{CS} + S_{\rm int}
	\label{eq:Lagrangian_continuum}
\end{equation}
where $S_{00}, S_{0j}$ and $S_{ij}$ account for  the temporal and spatial components of the polarization tensor. The $S_{CS}$ and $S_{\rm int}$ terms are obtained by taking the continuum limits of the Chern-Simons and interaction terms respectively. In obtaining these terms we make the further simplification that only terms to second order in derivatives are kept for the slowly fluctuating gauge fields. For the fast fluctuating gauge fields, only the leading order non-derivative terms are retained. The explicit expressions for the above terms are quite cumbersome and are saved for Appendix~\ref{app:Continuum_Action}. 

As mentioned above $S_{\rm final}$ has many more gauge fields than a usual gauge theory in $2+1$ dimensions. However, the important thing to note is that $S_{\rm final}$ is quadratic and massive in many of the fluctuating gauge fields. This will allow us to integrate out some of these extra fields and reduce the excessive number of gauge fields in this model. More precisely, the action in Eq.\eqref{eq:Lagrangian_continuum} is quadratic and massive in fields $A^{f 1}_0$, $A^{f 2}_0$, $A^f_1+A^f_2-A^f_3$ and $A_3$. In order to safely integrate out these fields, the mass terms have to have the correct sign i.e. the masses of the temporal fields must be positive and the masses of the spatial gauge fields must be negative. These conditions have been verified numerically for different values of $\lambda$. Once, these extra fields have been integrated out, we are just left with the traditional three fields $A_0, A_x$ and $A_y$ (after re-expressing the remaining fields along the $x$ and $y$ directions). (Recall that the above fields correspond to the fluctuating components and that we dropped the $\delta$ label to simplify the notation.)

The long-distance behavior of the final effective action can be  more succinctly expressed  as 
	\begin{align}
	S_{\rm eff}=&S_{CS}^{\rm eff}+S_M^{\rm eff}+\ldots \nonumber\\
S_{CS}^{\rm eff}= & \int d^3x \frac{1}{2}\left(\theta +\theta _F\right)\epsilon _{\mu \nu \lambda }A_{\mu }\partial _{\nu }A_{\lambda }\nonumber\\
S^{\rm eff}_{M}= & \int d^3x \left[ \frac{1}{2}\epsilon {\bm E}^2-\frac{1}{2}\chi  B^2 \right]
	\label{eq:Final_continuum_action}
\end{align}
where $S_M^{\rm eff}$ donates the Maxwell action with coefficients $\epsilon > 0$ and $\chi > 0$. $\bm E$ and $B$ are, respectively, the effective ``electric'' and ``magnetic'' fields of the statistical gauge field $A_\mu$.
(Please note that in two space dimensions the electric field is a vector while the magnetic field is a pseudoscalar.)

The effective Chern-Simons parameter for the $XY$ and Ising regimes are given in Table \ref{CSterm_final}.
In Eq.\eqref{eq:Final_continuum_action}, the labels $\mu, \nu$ now only take values $0, x, y$. To lowest order the most important term is just the effective Chern-Simons term $S_{CS}^{\rm eff}$ as it has the least number of derivatives. For $\lambda \lesssim 0.6$, the Chern-Simons term obtained by integrating out the fermions and the original Chern-Simons term obtained from the Jordan-Wigner transformation add where as, for $\lambda \gtrsim 0.6$, we are just left with the original Chern-Simons term.

\begin{table}
\begin{tabular}{|c|c|c|} \hline
 $\lambda$  & $\theta+\theta_F $\\ \hline 
	& \\
	$\lesssim 0.6$ & $\frac{1}{2\pi}  +\frac{1}{2\pi }$ \\
	&  \\
$ \gtrsim 0.6$ &$ \frac{1}{2\pi} + 0$  \\
	& \\
\hline
\end{tabular}
\caption{Effective Chern-Simons  parameter for the $XXZ$ model in the $XY$ and Ising regimes.}
	\label{CSterm_final}
\end{table}

The results of the above computation can be divided into two regimes. 
\subsubsection{XY regime}

In the $XY$-regime i.e. for $\lambda \lesssim 0.6$, the Chern-Simons terms add. The final low energy action (just keeping the Chern-Simons part) has the form\cite{Fradkin1990,Yang1993}  
\begin{equation}
	S_{\rm eff}^{CS}(A_{\mu}) = \frac{1}{2} \left( \theta_F+\theta \right)  S_{CS}(A_{\mu})
	\label{eq:CS_eff_final}
\end{equation}
Here $\theta_F$ is the coefficient of the induced Chern-Simons term obtained by integrating out the fermionic degrees of freedom and it is given by
\begin{equation}
\theta_F=\frac{C}{2\pi}
\label{eq:thetaF}
\end{equation}
and is the effective Hall conductivity of the mean field state.
Here $C$ is the total Chern number of the occupied bands of the mean field theory. 

In Eq.\eqref{eq:CS_eff_final} we have neglected terms in the effective action with more than one derivatives since they are irrelevant at long distances. As it is apparent from Eq.\eqref{eq:CS_eff_final},\cite{Fradkin1990,Yang1993}  
the physics of the full system (beyond mean field theory) depends on whether   $\theta+\theta_F$ vanishes or not.
In particular we will have a gapped state (with broken time reversal invariance) only if $\theta+\theta_F \neq 0$. Otherwise, if $\theta+\theta_F=0$, the Maxwell-type subleading terms control the low energy physics. In this case the system has a gapless excitation, a ``photon'', which is equivalent to a Goldstone boson of the antiferromagnet. (For a detailed discussion see Ref.[\onlinecite{Fradkin2013}].) Since we are working with $\theta=\frac{1}{2\pi}$, a gapless state will occur whenever $C=-1$.

The response of this system to an external field can be measured by introducing a small external gauge field $\widetilde{A}_{\mu}$. \cite{Lopez1991} In a FQH system this field is an external electromagnetic perturbation which induces a charge current. However, in the case of the antiferromagnet it is a field that induces an spin current of the form 
\begin{equation}
J_s({\bm r}, {\bm r}')=i (S_x({\bm r}) S_y({\bm r}')-S_y({\bm r}) S_x({\bm r}'))
\end{equation}
 at the link $({\bm x}, {\bm x}')$ of the Kagome lattice.
This  field $\widetilde{A}_{\mu}$ couples to the fermionic degrees of freedom in the same way as the statistical gauge fields $A_{\mu}$. In the presence of this perturbation the effective action becomes 
\begin{equation}
	S_{\rm eff}^{CS}(A_{\mu}, \widetilde{A}_{\mu}) = \frac{1}{2} \left( \theta_F S_{CS}(A_{\mu} +\widetilde{A}_{\mu}) + \theta  S_{CS}(A_{\mu}) \right)
\end{equation}
where again we have assumed that $\theta+\theta_F\neq 0$ and thus the Maxwell-type subleading terms can be safely ignored at low energies.

The actual response of the system to this perturbation is obtained by integrating out the statistical gauge fields. This gives
\begin{equation}
	S_{\rm eff}^{CS}(\widetilde{A}_{\mu}) = \frac{1}{2} \theta_{\rm eff} S_{CS}(\widetilde{A}_{\mu})
\end{equation}
where $\theta_{\rm eff}$ is given by
\begin{equation}
	\frac{1}{\theta_{\rm eff}} = \frac{1}{\theta}+\frac{1}{\theta_{F}} 
	\label{eq:theta_eff}
\end{equation}
This result yields a spin Hall conductance $\sigma_{xy}^s$ of
\begin{equation}
\sigma_{xy}^s=\frac{\theta_{\rm eff}}{2}
\label{eq:sigmaxy-s}
\end{equation}

\paragraph{The $\frac{1}{3}$ magnetization plateau:} Since  $\theta = \frac{1}{2\pi}$  and at the $\frac{1}{3}$ plateau we found $\theta_F = \frac{1}{2\pi}$ (from Table \ref{CSterm_final}), this implies that 
$\theta_{\rm eff} = \frac{1}{2} \frac{1}{2\pi}$.
Hence at the $\frac{1}{3}$ magnetization plateau Kagome antiferromagnet has a {\em fractional spin Hall conductivity}
\begin{equation}
\sigma^s_{xy} = \frac{1}{2}\frac{1}{2\pi}
\end{equation}
Hence, the fluctuation effects tell us that the ground state of the Kagome antiferromagnet in the $XY$ regime at $\frac{1}{3}$-filling resembles a $\nu=1/2$ Laughlin fractional quantum Hall   state for (hard-core) bosons. In fact, an alternative description of the state that used the hydrodynamic theory\cite{Wen-1995} shows that the effective field theory of this state is a level $k=2$ Chern-Simons gauge theory. The upshot of this analysis is then that the ground state of the Kagome spin-1/2 Heisenberg antiferromagnet at its magnetization plateau at $\frac{1}{3}$ is a topological fluid with a (spin) Hall conductivity of $\frac{1}{2} \frac{1}{2\pi}$, a {\em two-fold degenerate ground state} on a torus, {\em a single chiral edge state} (with compactification radius $\sqrt{2}$) on a disk, and excitations are semions with statistical angle $\pi/2$. The same results apply to the equivalent magnetization plateau at $\frac{2}{3}$.

\paragraph{The magnetization plateau at $\frac{5}{9}$:}
By extending this  analysis  to  the magnetization plateau at $\frac{5}{9}$, we obtain the results summarized in Table \ref{summary-XY}. The $\frac{5}{9}$ plateau has mean-field Chern number $C=2$, and hence it is equivalent to the first FQH daughter Jain state for bosons in the $\frac{1}{2}$ FQH bosonic Laughlin state. From standard results of the theory of the FQH states,\cite{Wen-1995} we can then predict that  the $\frac{5}{9}$ plateau has {\em two chiral edge states} on a disk and a {\em three-fold degenerate ground state} on the torus. In this case, the Hall spin conductivity is $\sigma^s_{xy}=\frac{2}{3} \frac{1}{2\pi}$ and the excitations are anyons with statistical angle $\pi/3$.
\begin{table}[hbt]
	\begin{tabular}{|c|c|c|c|c|} \hline
	$\langle n \rangle$ & $m$ & $2\pi\theta_{F}$  & $\sigma_{xy}$ & $\delta$ \\ \hline
	& & & & \\
	$\frac{1}{6}$ & $\frac{2}{3}$ & $+1$ & $\frac{1}{2}$ & $\frac{\pi}{2}$\\
	& & & & \\
	$\frac{2}{9}$ & $\frac{5}{9}$ & $+2$ & $\frac{2}{3}$ & $\frac{\pi}{3}$\\
	& & & & \\
	$\frac{1}{3}$ & $\frac{1}{3}$ & $+1$ & $\frac{1}{2}$ & $\frac{\pi}{2}$\\
	& & & & \\	
 \hline
	\end{tabular}
	\caption{Summary of results for the magnetization plateaus of the $XY$ model at $1/3$, $5/9$ and $2/3$.}
	\label{summary-XY}
\end{table}

At all other fillings, the expected value of $\theta_F$ is $\theta_{F} = -\frac{1}{2\pi}$. In these regions the pre-factor of the effective Chern-Simons term in Eq.\eqref{eq:CS_eff_final}   exactly cancels out, i.e. $\theta + \theta_{F} = \frac{1}{2\pi}\left(1+ (-1) \right)=0$, leaving just the Maxwell term in Eq.\eqref{eq:Final_continuum_action} as the leading term. In this case, the elementary excitation is not a vortex but a ``photon''. However, in $2+1$ dimensions there is only one possible polarization state for a photon and this problem then turns out to be equivalent to a system with a Goldstone boson. In other words, away from the plateaus the transverse excitations are Goldstone modes of the spontaneously broken residual $U(1)$ symmetry.
This is also the behavior that one expects in the low density (low filling $\langle n \rangle \to 0$) regime when the lattice is sparsely filled and frustration effects are minimal.

Finally we consider the implications if we had chosen $\theta = -\frac{1}{2\pi}$, instead, in our original Jordan-Wigner transformation. As noted in the introduction as well as in our discussion on the mean-field physics, flipping the sign of $\theta$ breaks the time-reversal symmetry in the opposite manner and yields a new set of degenerate states. Again, at the mean-field level this gave rise to an integer quantum Hall state with the opposite Chern number $C = -1$. When the effect of fluctuations are taken into about, this now corresponds to an effective Chern-Simons term with $\theta + \theta_F = -\frac{1}{\pi}$. This again describes a fractional quantum Hall state for bosons but with opposite spin Hall conductivity $\sigma^s_{xy} = -\frac{1}{2}$.

\subsubsection{Ising regime}
\label{sec:Ising}

In the Ising regime ($\lambda \gtrsim 0.6$), $\theta_F = 0$. Hence, the effective Chern-Simons theory just has the parameter $\theta$ from the Jordan-Winger transformation and the statistical angle is simply $\delta = \pi$. This effectively transmutes the fermions back to the bosons that we began with. A similar analysis has been performed in the case of the square lattice by L\'opez et. al. \cite{Lopez1994} and they obtained a similar result in the Ising regime. The square lattice is unfrustrated and the spins behaving like bosons leads to the familiar N{\'e}el state on the square lattice in the Ising regime. However, the Kagome lattice is still frustrated in the Ising limit and our analysis does not choose a specific configuration. An analysis of quantum order by disorder is the needed.
There is a lot of numerical evidence that indicates that in the Ising regime, the Kagome lattice favors a VBC type state with a much larger unit cell (based on a $\sqrt{3} \times \sqrt{3}$ structure),\cite{Cabra2005, Capponi2013, Nishimoto2013} and possibly a $\mathbb{Z}_2$ time-reversal-invariant topological phase in between as in the simpler Ising systems.\cite{Moessner-2001b}

\section{Spin correlations in the Magnetization Plateaus}
\label{sec:spin-correlations}

\subsection{$S^zS^z$ correlations}
\label{sec:SzSz-correlations} 

The fluctuating component of the $\langle S^z S^z \rangle$ correlation can be computed by expressing them in terms of the magnetic fields as follows
\begin{equation}
	\langle S^z (x) S^z(y) \rangle = \frac{1}{4}+ \theta ^2 \langle B (x) B (y) \rangle
\end{equation}
This calculation can be performed by introducing the usual source term $j_{\mu}$ coupled to a Chern-Simons term. The details are presented in Appendix~\ref{app:SzSz_expectation}. The Fourier transform of the connected $S^zS^z$ spin correlation, i.e. the longitudinal magnetic susceptibility at momentum $\bm p$ and frequency $\omega$ and which we denote by $\chi^{zz}(p)$ (here $p=(\bm p,\omega)$), can be expressed as follows
	\begin{align}
	\chi^{zz}(p) = i  \theta^2 & \Big( p_1^2 G_{22}(p) -p_1 p_2 G_{12}(p)\nonumber \\
						 &-p_1 p_2 G_{21}(p)+p_2^2 G_{11}(p)\Big)
	\end{align}
where $G_{\mu\nu}(p)$ is the Fourier transform of the propagator of the gauge field  of the continuum action given in  Eq.\eqref{eq:Final_continuum_action}.  This propagator  can easily be computed by introducing a usual gauge fixing term. In the Lorentz gauge $\partial^{\mu}A_{\mu} = 0$. This yields the following  value for the longitudinal (time-ordered) magnetic susceptibility  in the low energy and long wavelength limit
\begin{equation}
\chi^{zz}(p) = \theta^2 \frac{\epsilon {\bm p}^2}{ \epsilon^2 \omega^2 - \epsilon \chi {\bm p}^2 - (\theta + \theta_{F})^2+i\eta}
\end{equation}
where ${\bm p}^2 = p_1^2 + p_2^2$, and as usual we must take the limit $\eta \to 0^+$. Hence, in the magnetization plateaus the longitudinal magnetic susceptibility has a Lorentzian shape centered at zero momentum with a  width determined by the energy gap $\sim (\theta+\theta_F)/\epsilon$.

For the case of the $m = \frac{1}{3}$ plateau $\theta + \theta_F = \frac{1}{\pi}$ and the collective modes are gapped. At all other fillings on the Kagome lattice that do not correspond to these plateau type states or for the case of the square lattice where $\theta + \theta_F = 0$, the collective modes would be gapless and correspond to Goldstone modes of the transverse fluctuations.

\subsection{$XY$ correlations} 
\label{sec:XY-correlations}

The transverse, $XY$, components  can be expressed as 

	\begin{align}
	S^x(x)S^x(y) + S^y(x)S^y(y) & =\nonumber \\
			 \frac{1}{2}  &\left( S^+(x)S^-(y) + S^+(y)S^-(x) \right)
	\end{align}
Hence, the computation of the above expectation value boils down to a computation of $\langle S^+(x)S^-(y) \rangle$. From the Jordan-Wigner transformation discussed in subsection \ref{sec:JW}, we know that this can be expressed in the continuum as (plus its Hermitian conjugate which is not written down explicitly)
	\begin{align}
		S^{XY}_{\Gamma}(x,y) = & \langle S^+(x)S^-(y)\rangle \nonumber \\ 
					= & \langle\psi^{\dagger}(x) e^{i \int_{\Gamma(x,y)} A_{\mu}dx^{\mu}} \psi(y) \rangle \\
			= & \langle G_F(x,y;A_{\mu}) e^{i \int_{\Gamma(x,y)} A_{\mu}dx^{\mu}} \rangle_{A_{\mu}}
	\end{align}
where $G_F(x,y; A_{\mu})$ is the fermion propagator  in the presence of the statistical gauge field $A_{\mu}$. In the second step we have just averaged over the fermionic degrees of freedom. This gives rise to the expectation of the operator in the last step where the average is just over the statistical gauge fields $A_{\mu}$. 

This expectation is quite difficult to compute in general due to the non-local string along the path $\Gamma(x,y)$ that connects points $x$ and $y$. To simplify the above expression, the fermionic propagator can be expressed in terms of a Feynman path integral as a sum over histories of the particles. At a semi-classical level the above integral can then be expressed in terms of smooth trajectories which are the dominant contributions for a problem with an energy gap and in the long-distance limit. The below simplifications also rely on the fact that the photon propagator is massive as in the case of the $m= \frac{1}{3}$ plateau. If this were not the case the below results would be drastically altered. The details of this calculation are shown in Appendix~\ref{app:XY_correlations}.

Finally, the transverse correlation (at equal times) can be approximated as
\begin{equation}
	S^{XY}_{\Gamma}(x,y) \simeq \sum_{\gamma}({\rm Amp})_{\gamma} e^{I_1 + I_2}
\end{equation}
where the $({\rm Amp})_{\gamma}$ is the weight associated with a smooth trajectory ${\gamma}$. The set of paths $\{\gamma\}$ are closed curves which are the oriented sum of the paths $\Gamma$ and its histories.

The first integral $I_1$ in the exponent is proportional to the length of the path associated with the curve $\gamma$ and is given by
	\begin{equation}
		I_1 =  -\frac{\epsilon}{2 (\theta+\theta_F)^2} L(\gamma)
\end{equation}
where $L(\gamma)$ is the perimeter of the path $\gamma$. 

The second integral $I_2$ can be expressed as 
\begin{equation}
		I_2 \approx \frac{i}{2\bar{\theta}} \frac{\Phi_{\gamma}}{\phi_0} = \frac{i}{2} \Phi_{\gamma}
\end{equation}
where $\Phi_{\gamma}$ is the average flux over the path $\gamma$. Hence, the term $e^{I_2}$ corresponds to an Aharonov-Bohm phase associated with the path $\gamma$. In the last step, we have used the fact that $\bar{\theta} = \theta + \theta_F = \frac{1}{\pi}$ and the flux quantum $\phi_0 = 2\pi$.

If one chooses the path $\gamma$ to correspond to the shortest path between the points $x$ and $y$ (i.e. the path that minimizes the classical equations of motion), then the expectation can be further simplified as 
\begin{equation}
	S^{XY}_{\Gamma}(x,y) \simeq \sum_{s.p} ({\rm Amp})_{s.p.} \exp\left\{-\frac{\epsilon}{2\bar{\theta}^2 }L_{s.p.}+\frac{i}{2}\Phi_{s.p.} \right\}
\end{equation}
where $s.p.$ correspond to the classically shortest paths and $\bar{\theta} = \theta + \theta_F$ again, and $\epsilon$ is defined in the effective action of Eq.\eqref{eq:Final_continuum_action}. Explicit numerical values of $\epsilon$ are given in Table \ref{table:coefficients} of Appendix \ref{app:Continuum_Action}

\paragraph{Case 1:} In this first case, the trajectory of the classically shortest path does not enclose any loops. As an example one could consider measuring this $XY$ correlation between a point $x$ and another point along $e_1$ direction of the lattice (in Fig.\ref{fig:Kagome_unit_cell}) at say $x + d_x$. In this case there is just one classical path corresponding to the straight line between points $x$ and $y$ and the area of the loop associated with the Aharonov-Bohm phase would reduce to zero. In this case, the above equal-time transverse correlation function would simply reduce to  
\begin{equation}
	S^{XY}_{\Gamma}(x,x + d_x) \approx  e^{-\epsilon \pi^2 d_x}
\end{equation}
i.e. just exponentially decaying in the distance between the two points. Glossing over subtleties, the path between arbitrary points $\bm x$ and $\bm y$ can always be chosen so that the correlation function decays exponentially. Then, from the above result we can infer that the correlation length is $\xi \simeq 1/(\pi^2 \epsilon)$, where the values of $\epsilon$ are given in Table \ref{table:coefficients} of Appendix \ref{app:Continuum_Action}.  In Fourier space this transverse correlation function is a Lorentzian.

\paragraph{Case 2:} In the second case, the classical trajectory can enclose some number of hexagons leading to an Aharonov-Bohm phase factor. For instance, let us consider the correlation between the point $x$ and another point, say, along the $e_1 + e_2$ direction of the lattice in Fig.\ref{fig:Kagome_unit_cell}. This would correspond to points along the diagonal of the hexagons. Classically there are four paths/ways to reach the point across the diagonal of the hexagon, two of which lead to a phase of $\Phi_{\rm hex} = \frac{2\pi}{3}$ (in the $XY$ limit). Hence, the $XY$ correlation picks up an additional phase of $(2+2 e^{i \frac{2\pi}{3}})$ for each hexagon that the classical trajectory encounters. More generally, for a path that encloses a certain number $n$ of hexagons and the generic result would be
	\begin{align}
	S^{XY}_{\Gamma}(x, y) & \simeq  2^n(1+ e^{i \frac{2\pi}{3}})^n e^{-\epsilon \pi^2 d_{(x,y)}}\nonumber \\
	S^{XY}_{\Gamma}(x, y) & \simeq  2^n e^{i n\frac{\pi}{3}} e^{-\epsilon \pi^2 d_{(x,y)}} 
	\end{align}
where $d_{(x,y)}$ is the distance along a classical trajectory from $x$ to $y$ and then back. Here there are several different Aharonov-Bohm phases that can arise depending on the number of hexagons encircled by a classical trajectory (The distinct values would correspond to $n = 0,...,5$).

\section{Conclusion and future work}
\label{sec:conclusions}

We have analyzed the nearest-neighbor spin-1/2 Heisenberg antiferromagnet on the Kagome lattice using a two dimensional Jordan-Wigner transformation that maps spins to fermions coupled to a Chern-Simons gauge field. This transformation relies on writing down a lattice Chern-Simons term on the Kagome lattice which was recently developed and outlined here.

We  used this model to look at magnetization plateaus that can arise at the mean-field level. At the mean-field level these states had integer Chern numbers. Hence, at the mean-field level these plateaus corresponded to integer quantum Hall type states. In the case of $\frac{1}{3}$-filling ($m = \frac{1}{3}$ plateau) we found that in the $XY$-regime, the inclusion of fluctuations gives rise to an effective Chern-Simons theory that predicts a fractional quantum Hall state of bosons with a spin Hall conductivity of $\sigma^s_{xy} = \frac{1}{2} \frac{1}{2\pi}$, thus altering the mean-field physics drastically. The excitations of this plateau are anyons with fractional spin and statistical angle $\frac{\pi}{2}$ (i.e. are semions). This plateau state is two-fold degenerate on a torus. On a disk geometry it supports a chiral edge state (with compactification radius $\sqrt{2}$).
In the Ising regime, the effective Chern-Simons theory just transmutes the fermions back to the original bosons and reduces to earlier results. We also extended  this analysis to the case of the $\frac{2}{3}$ and $\frac{5}{9}$ magnetization plateaus. The $\frac{2}{3}$ plateau is essentially equivalent to the $\frac{1}{3}$ plateau. However, the $\frac{5}{9}$ plateau turns out to be equivalent to the first Jain state for bosons with (spin) Hall conductance of $\sigma^s_{xy}=\frac{2}{3} \frac{1}{2\pi}$.  Its excitations are anyons with statistical angle $\frac{\pi}{3}$,
This state has a three-fold ground state degeneracy on a torus,  and supports two chiral chiral gapless  edge states on a disk geometry. Due to the gapless edge states of  both the $\frac{1}{3}$ and in the $\frac{5}{9}$ plateaus, there should give a linear temperature-dependent contribution to the heat capacity. 
Likewise there should be a universal finite thermal conductivity in the plateau states due to the edge states.

In the Introduction we noted that the magnetization plateaus of the spin-1/2 quantum Heisenberg model had been investigated by Cabra and coworkers.\cite{Cabra2005}
These authors used exact diagonalizations of small systems (with up to 36 spins) to investigate the behavior of the 1/3 plateau for a wide range of anisotropies,  from the Ising to the $XY$ regimes, and compared their results with semiclassical ($1/S$) expansions and with effective quantum-dimer-like Hamiltonians derived in the Ising regime. In their work they did not see any hint of a phase transition as the anisotropy changed from Ising type to $XY$ like and hence their small system diagonalizations do not seem to show evidence for a chiral spin liquid state in the $XY$regime. However, even if the chiral spin liquid phase were present,  in such small systems (containing a total of twelve Kagome unit cells) the degeneracy of the two topologically inequivalent states should be lifted.  Unfortunately the methods that we use in this work provide a reliable analysis of the long distance (and low energy) behavior of the system but do not provide a reliable estimate of the value of the energy gap, needed to make a meaningful comparison with the numerical results of Ref. [\onlinecite{Cabra2005}]. It is likely that more sophisticated methods, such as DMRG and its tensor network generalizations, should be able to provide evidence for such states.

Finally, we would like to end with a few remarks on the half-filled case (no external field) which was not considered here. The magnetization plateau type states that were considered here were gapped (even at the mean-field level). However at half-filling, our theory predicts a gapless state with Dirac points and a doubled unit cell at the mean-field level. This state looks very similar to the $U(1)$-Dirac spin liquid discussed in other works.\cite{Ran2007,Iqbal2011} A discussion on this state and the effect of fluctuations will be saved for a future work.

In a recent paper, Bauer and coworkers\cite{Bauer-2014} studied a spin-1/2 Heisenberg antiferromagnet on the kagome lattice at zero external magnetic field (i.e. the half-filled case). These authors also include a coupling to the spin chirality operator on each triangle, a term that breaks time-reversal invariance explicitly. The main result of their work is that, apparently for all non-vanishing values of the chiral coupling constant, the ground state of the system is a chiral spin liquid state which is equivalent to a Laughlin fractional quantum Hall state   for bosons at filling fraction $\nu=1/2$. This is in fact the same state that we find here in the 1/3 plateau state at least with $XY$ anisotropy. It is tempting to speculate that these two regimes may be somehow connected to each other. In particular, their numerical results for the entanglement properties apply to the chiral spin liquid state we found for the 1/3 (and 2/3) plateaus since they have the same universal effective field theory.

\begin{acknowledgments}
We thank Bryan Clark, Hitesh Changlani, Greg MacDougall, and Michael Lawler for useful discussions. This work was supported in part by the National Science Foundation, under grants DMR-1064319 and DMR-1408713 at the University of Illinois and PHY-1402971 at the University of Michigan, and by the U.S. Department of Energy, Division of Materials Sciences under Awards No. 
DE-FG02-07ER46453 and  DE-Sc0012368 through the Frederick
Seitz Materials Research Laboratory of the University of Illinois at Urbana-Champaign.
\end{acknowledgments}

\appendix
\begin{widetext}
\section{Continuum Action}
\label{app:Continuum_Action}

This section of the appendix presents the details on obtaining the full continuum action in Sec \ref{sec:Full_continuum_action}. We begin by computing the components of the polarization tensor (corresponding to all the different fluctuating components shown in Sec \ref{sec:Full_continuum_action}) in the long wavelength limit. Then we integrate out the fast components and obtain an effective low energy action containing only the slowly fluctuating components.

The polarization tensor is most easily computed in momentum space using the below expression
\begin{equation}
	\Pi_{\mu\nu}(P_0, \vec{P}) = \int_{\omega, \vec{q}} i Tr\left[S\left(P_0+\omega ,\vec{P}+\vec{q}\right)j_{\mu} (\vec{P}+\vec{q})S(\omega ,\vec{q}) j_{\nu}(\vec{q}) \right]
	\label{eq:Polarization_tensor}
\end{equation}
where $S(\omega,\vec{q})$ is the mean-field fermion propagator as shown in Eq.\eqref{eq:MF_propagator} and $j_{\mu}$ are the currents of the corresponding fluctuating components. The currents $j_{\mu}$ can be computed by taking derivatives of the action with respect to the corresponding fluctuating components i.e. $j_{\mu} = - \frac{\delta S}{\delta A_{\mu}}$. It is in general challenging to evaluate the integral shown in Eq.\eqref{eq:Polarization_tensor} due to the complicated form of the propagator in Eq.\eqref{eq:MF_propagator}. To simplify the calculation, we focus on the long wavelength limit by expanding Eq.\eqref{eq:Polarization_tensor} in powers of momenta $P$ (up to quadratic order). This will allow us to compute the above integrals numerically for each of the fluctuating components in Eq.\eqref{eq:Temporal_fluctuations} and Eq.\eqref{eq:Spatial_fluctuations}. As the slow components are considered to be more relevant, we make the following simplifications. Firstly, for the slowly fluctuating components, we only keep terms to quadratic order in momenta. Secondly, for the fast fluctuating components, we only keep the leading order terms i.e. the non-derivative terms. 
The continuum limit of the polarization tensor will then have a bunch of non-derivative (or mass) terms to the leading order. The remaining derivative terms (only for the slow components) takes the following form  
	\begin{align}
A_0(P)\Pi ^{00}A_0(-P)= &  u_{0xx0} A_0(P) P_x^2 A_0(-P)+u_{0yy0} A_0(P) P_y^2 A_0(-P)\nonumber \\
A_0(P)\Pi ^{0j}A_j(-P)= & u_{0xj}A_0(P)i P_xA_j(-P)+u_{0yj}A_0(P)i P_yA_j(-P)\nonumber \\
							& +u_{0oxj}A_0(P)\Omega  P_xA_j(-P)+u_{0oyj}A_0(P)\Omega  P_yA_j(-P)\nonumber \\
A_i(P)\Pi ^{ij}A_j(-P)= & u_{ioj}A_i(P)i \Omega  A_j(-P)+ u_{iooj}A_i(P)\Omega ^2A_j(-P)	\nonumber \\
								& +u_{ixxj} A_i(P)P_x^2A_j(-P)+u_{iyyj}A_i(P)P_y^2A_j(-P)+u_{ixyj}A_i(P)P_xP_yA_j(-P)
								\label{eq:Pimunu-expansion}
	\end{align}
where $\Omega$ is the time component of the three-momentum $P_\mu$ (with $\mu=0,x,y$), and $P_x$ and $P_y$ refer to the spatial components of the three-momentum $P_\mu$.
Also note that all the fields above correspond to the fluctuating components (we have dropped the $\delta$ label to make the notation less cumbersome). 
The $u$ coefficients are obtained by performing the integral in Eq.\eqref{eq:Polarization_tensor}. For example the coefficients  $u_{0xx0}$ and $u_{0yy0}$ in Eq.\eqref{eq:Pimunu-expansion} correspond to the $\Pi_{00}$ component when expanded out in powers of momenta $P$ (up to second order).
This calculation can be further simplified by realizing that the transversality condition puts some constraints on these $u$ coefficients. The action can then be written as
\begin{equation}
S_{\rm final} = S_{00} + S_{0j} + S_{ij} + S_{CS}^1 + S_{CS}^2 + S_{int,0} + S_{int,2}
	\label{eq:Lagrangian_continuum2}
\end{equation}
where the corresponding Lagrangian densities are given by
\begin{equation}
\mathscr{L}_{00}= \frac{1}{2}\, u^{fi fi}_{00} \left[\left(A^{f 1}_0\right)^2+\left(A^{f 2}_0\right)^2\right]-\frac{1}{2}u^{f1f2}_{00} A^{f 1}_0A^{f 2}_0-\frac{1}{2}u_{0xx0}A_0 \partial _x^2 A_0-\frac{1}{2}u_{0yy0}A_0\partial _y^2 A_0
\end{equation}
\begin{equation}
	\begin{aligned}
\mathscr{L}_{0j}= & \frac{1}{2}u^{fi f}_{0j}\left\{-\left[A^{f 1}_0 +A^{f 2}_0\right]A^f_1-\left[A^{f 1}_0 +A^{f 2}_0\right] A^f_2+\left[A^{f1}_0 +A^{f 2}_0\right] A^f_3\right\} \\
		& - A_0\partial _x\left[u_{0x1}A_1+u_{0x2}A_2+u_{0x3}A_3\right]-A_0\partial
_y\left[u_{0y1}A_1+u_{0y2}A_2+u_{0y3}A_3\right] \\ 
		& +A_0\partial _0\partial_x\left\{u_{0ox1}A_1+u_{0ox2}A_2+u_{0ox3}A_3\right\}+A_0\partial
_0\partial _y\left\{u_{0oy1}A_1+u_{0oy2}A_2+u_{0oy3}A_3\right\}
	\end{aligned}
\end{equation}
\begin{equation}
	\begin{aligned}
\mathscr{L}_{i j}=\frac{1}{2} & u^{ff}_{jk}\left(A^f_1+A^f_2-A^f_3\right)^2+ \frac{1}{2}u_{jk}\left(A_1+A_2-A_3\right)^2 \\
		+& \left\{u_{1o2}A_1\partial_0A_2+u_{1o3}A_1\partial _0A_3+u_{2o3}A_2\partial _0A_3\right\} \\ 
		-& \frac{1}{2}\left\{u_{1oo1}A_1\partial _0^2A_1+u_{2oo2}A_2\partial
_0^2A_2+u_{3oo3}A_3\partial _0^2A_3+2u_{1oo2}A_1\partial _0^2A_2+2u_{1oo3}A_1\partial _0^2A_3+2u_{2oo3}A_2\partial_0^2A_3\right\} \\
		-&\frac{1}{2}\left\{u_{1xx1}A_1\partial_x^2A_1+u_{2xx2}A_2\partial_x^2A_2
		+u_{3xx3}A_3\partial _x^2A_3+2u_{1xx2}A_1\partial_x^2A_2
		+2u_{1xx3}A_1\partial _x^2A_3+2u_{2xx3}A_2\partial _x^2A_3\right\} \\
		-&\frac{1}{2}\left\{u_{1yy1}A_1\partial_y^2A_1+u_{2yy2}A_2\partial _y^2A_2+u_{3yy3}A_3\partial _y^2A_3+2u_{1yy2}A_1\partial _y^2A_2+2u_{1yy3}A_1\partial
_y^2A_3+2u_{2yy3}A_2\partial _y^2A_3\right\} \\
		-&\frac{1}{2}\big\{u_{1xy1}A_1\partial _x\partial _yA_1+u_{2xy2}A_2\partial
_x\partial _yA_2+u_{3xy3}A_3\partial _x\partial _yA_3 \\
			&+2u_{1xy2}A_1\partial _x\partial _yA_2+2u_{1xy3}A_1\partial _x\partial_yA_3+2u_{2xy3}A_2\partial _x\partial _yA_3\big\}
	\end{aligned}
\end{equation}
\begin{equation}
	\begin{aligned}
\mathscr{L}^1_{CS}= & \frac{4}{\sqrt{3}a_0} \theta  \left\{\left(A^{f 1}_0+A^{f 2}_0\right) \left[A^f_1+A^f_2-A^f_3\right]+\left(-A^{f 1}_0+A^{f2}_0\right)\left[A_1+A_2-A_3\right]\right\} \\
						& +\frac{1}{2\sqrt{3}}\theta  A_0 \partial _x\left(A_1+5 A_2+3 A_3\right)+\frac{1}{2}\theta  A_0\partial_y\left(-3 A_1+A_2-A_3\right) \\
\mathscr{L}^2_{CS}= & \frac{2}{\sqrt{3} } \theta  \left\{A_2 \left(2 \partial _0A_1+\partial _0A_3\right)-A_1 \partial _0A_3\right\}
		\end{aligned}
\end{equation}
\begin{equation}
	\begin{aligned}
\mathscr{L}_{int,0}= & \frac{4}{\sqrt{3} a_0} J \theta ^2 \lambda  \left\{3 \left(A^f_1+A^f_2-A^f_3\right)^2+\left(A_1+A_2-A_3\right)^2\right\}\\
\mathscr{L}_{int,2}= -J \lambda  \theta ^2\frac{2}{\sqrt{3} } a_0\big\{ & \left(-\partial _1A_1-2\partial _2A_1+2\partial _1A_2+\partial _2A_2+\partial_1A_3-\partial _2A_3\right)\left[-2\partial _2A_1+2\partial _1A_2\right] \\ 
					& -\left(A_1+A_2-A_3\right)\partial _3^2\left(A_1+A_2-A_3\right)\big\}
	\end{aligned}
\end{equation}
where $a_0$ is the lattice spacing and the continuum limit amounts to taking the limit $a_0 \to 0$. $\mathscr{L}_{00}$, $\mathscr{L}_{0j}$ and $\mathscr{L}_{ij}$ are obtained from the polarization tensor in Eq.\eqref{eq:Polarization_tensor} using the procedure outlined above. The $\mathscr{L}_{CS}$ and $\mathscr{L}_{\rm int}$ terms are obtained by expressing the Chern-Simons term in Eq.\eqref{eq:Generic_CS_term} and interaction term Eq.\eqref{eq:Interaction_A} in terms of the fluctuation components and taking the continuum limits respectively. 
The leading order mass terms are all proportional to $\frac{1}{a_0}$.

As noted in sections~\ref{sec:Effects_of_fluctuations} and ~\ref{sec:Full_continuum_action}, the above action has excessive gauge fields. However, many of these fields are quadratic and massive and can be integrated out. More precisely, the above action has massive fields corresponding to $A_0^{f1}$, $A_0^{f2}$, $\widetilde{A}_3 \equiv A_1 + A_2 -A_3$ and $\widetilde{A}_3^f \equiv A_1^f + A_2^f - A_3^f$.  Re-writing the quadratic part of the action that will be integrated out gives
\begin{equation}
\mathscr{L}_{quad}=  \frac{1}{2}A^T\cdot M \cdot A+A \cdot N 
\end{equation}
with 
\begin{equation}
A =\left(
\begin{array}{c}
 A^{f 1}_0 \\
 A^{f 2}_0 \\
  \tilde{A}^f_3 \\
 \tilde{A}_3 \\
\end{array}
\right),\quad 
N=\left(
\begin{array}{ccc}
 0 & 0 & 0 \\
 0 & 0 & 0 \\
 0 & 0 & 0 \\
 \Delta _{30} & \Delta _{31} & \Delta _{32} \\
\end{array} 
\right) 
\left(
\begin{array}{c}
A_0 \\ A_1 \\ A_2
\end{array}
\right) 
\end{equation}
and 
\begin{equation}
M=\left(
\begin{array}{cccc}
 \, u^{fi fi}_{00}  & -\frac{1}{2}u^{f1 f2}_{00}  & \frac{1}{2}u^{fi f}_{0j}-\frac{4}{\sqrt{3}a_0} \theta
 & \frac{4}{\sqrt{3}a_0} \theta  \\
 -\frac{1}{2}u^{f1 f2}_{00}  & \, u^{fi fi}_{00}  & \frac{1}{2}u^{fi f}_{0j}-\frac{4}{\sqrt{3}a_0} \theta
 & -\frac{4}{\sqrt{3}a_0} \theta  \\
 \frac{1}{2}u^{fi f}_{0j}-\frac{4}{\sqrt{3}a_0} \theta  & \frac{1}{2}u^{fi f}_{0j}-\frac{4}{\sqrt{3}a_0} \theta  & u^{ff}_{jk}+\frac{24}{\sqrt{3}
a_0} \theta ^2 \lambda   & 0 \\
 \frac{4}{\sqrt{3}a_0} \theta  & -\frac{4}{\sqrt{3}a_0} \theta  & 0 & u_{jk}+\frac{8}{\sqrt{3} a_0} \theta ^2 \lambda +2\Delta ^2 \\
\end{array}
\right)
\end{equation}
The $\Delta$'s correspond to derivative terms and are shown below.
\begin{equation}
	\begin{aligned}
\Delta ^2= & \lambda  \theta ^2\frac{2}{\sqrt{3} } a_0\partial _3^2-\frac{1}{2}\left\{u_{3oo3}\partial _0^2 +u_{3xx3}\partial _x^2+u_{3yy3}\partial
_y^2+u_{3xy3}\partial _x\partial _y\right\}\\
\Delta _{30}= & -\frac{3}{2\sqrt{3}}\theta  \partial _x+\frac{1}{2}\theta  \partial _y+ u_{0x3}\partial _x+ u_{0y3} \partial _y+u_{0ox3}\partial_0\partial _x+u_{0oy3}\partial _0\partial _y\\
\Delta _{31}= & -u_{1o3}\partial _0+\frac{2}{\sqrt{3} } \theta  \partial _0- \lambda  \theta ^2\frac{4}{\sqrt{3} } a_0\partial _1\partial _2A_1+\lambda  \theta ^2\frac{4}{\sqrt{3} } a_0\partial _2^2  \\
	& - \left\{u_{3oo3}\partial _0^2+u_{1oo3}\partial_0^2+u_{3xx3}\partial_x^2+u_{1xx3}\partial_x^2  +u_{3yy3}\partial _y^2+u_{1yy3}\partial _y^2
			+u_{3xy3}\partial _x\partial_y+u_{1xy3}\partial
_x\partial _y\right\}\\
\Delta _{32}= & -u_{2o3}\partial _0-\frac{2}{\sqrt{3} } \theta\partial _0+ \lambda  \theta ^2\frac{4}{\sqrt{3} } a_0\partial _1^2-\lambda  \theta ^2\frac{4}{\sqrt{3} } a_0\partial _1\partial_2 \\
				& -\left\{u_{3oo3}\partial _0^2+u_{2oo3}\partial _0^2+u_{3xx3}\partial
_x^2+u_{2xx3}\partial _x^2+u_{3yy3}\partial _y^2+u_{2yy3}\partial _y^2+u_{3xy3}\partial _x\partial _y+u_{2xy3}\partial
_x\partial _y\right\}
	\end{aligned}
\end{equation}
The sign of the masses have been verified numerically for different values of $\lambda$ and we have ensured that they have the correct sign in order to safely integrate them out. After integrating out these gaussian fields, the remaining terms are given by
\begin{equation}
\mathscr{L}_{\rm eff}=-\frac{1}{2}N^T \cdot M^{-1}\cdot N 
\end{equation}
To quadratic order, we find that
\begin{equation}
	\begin{aligned}
\mathscr{L}_{CS,eff}= & \frac{1}{2}\left(\theta +\theta _F\right)\epsilon _{\mu \nu \lambda }A_{\mu }\partial _{\nu }A_{\lambda } \\
\mathscr{L}_{EM}= & \frac{1}{2}\pi _{xx}\left[-A_0 \partial _x^2 A_0+2A_0\partial _0\partial _xA_x-A_x\partial _0^2A_x\right] +\frac{1}{2}\pi_{yy}\left[-A_0\partial _y^2A_0+2A_0\partial _0\partial _yA_y-A_y\partial _0^2A_y\right]\\
		&  +\frac{1}{2}\pi _{xy}\left[-A_y\partial _x^2A_y+2A_x\partial _x\partial _yA_y-A_x\partial _y^2A_x\right] - \lambda  \theta ^2\frac{3\sqrt{3}}{16}a_0\left(\partial_xA_y-\partial _yA_x\right)^2
	\end{aligned}
\end{equation}
with
\begin{equation}
	\begin{aligned}
	\theta _F= & -\frac{\sqrt{3}}{8}\left(u_{1o2}+u_{1o3}-u_{2o3}\right) \\
	\pi _{xx}= & \frac{1}{4}u_{1oo1}+\frac{1}{2}u_{1oo3}-\frac{1}{16}u_{2oo2}-\frac{1}{8}u_{2oo3}+\frac{3}{16}u_{3oo3}\\
	\pi _{yy}= & \frac{3}{16}\left(u_{3oo3}+2 u_{2oo3}+ u_{2oo2}\right) \\
	\pi _{xy}= & \frac{3}{16}\left(u_{2xx2}+2u_{2xx3}+u_{3xx3}\right)
	\end{aligned}
\end{equation}
We have also re-expressed the $A_1$ and $A_2$ fields along the $x$ and $y$ directions to give $A_x$ and $A_y$. Once again, all the $u$ coefficients are computed using the equation for the polarization tensor in Eq.\eqref{eq:Polarization_tensor}. The final form is written in terms of $E$ and $B$ fields as
\begin{equation}
	\begin{aligned}
\mathscr{L}_{CS,eff}= & \frac{1}{2}\left(\theta +\theta _F\right)\epsilon _{\mu \nu \lambda }A_{\mu }\partial _{\nu }A_{\lambda }\\
\mathscr{L}_{EM}= & \frac{1}{2}\epsilon {\bm E}^2- \frac{1}{2}\chi B^2
	\end{aligned}
	\label{eq:Final_action_app}
\end{equation}
where $\epsilon = \pi_{xx} = \pi_{yy}$, $\chi =g-\pi _{xy}$ and $g=\lambda  \theta ^2\frac{3\sqrt{3}}{8}a_0$ corresponding the interaction strength. Eq.\eqref{eq:Final_action_app} is the same result that was expressed in Section\ref{sec:Full_continuum_action}. 

The numerically computed values are listed in the below table for a few different values of $\lambda$.
\begin{table}
\begin{tabular}{|c||c|c|c|c|} \hline
 $\lambda$  &  $\theta _F+\theta$  & $\epsilon =\pi _{xx}=\pi _{yy}$ & $\chi =g-\pi _{xy}$ & $\pi _{xy}$ \\ 
\hline 
& & & & \\
 0 & $\frac{1}{2\pi }(2)$ & $\frac{a_0}{2\pi }1.33312$\, \,  & $\frac{a_0}{2\pi }0.242483$\,  & $-\frac{a_0}{2\pi }0.242466$ \\
& & & & \\
 0.25 & $\frac{1}{2\pi }(2)$ & $\frac{a_0}{2\pi }1.11182$\,  & $\frac{a_0}{2\pi }0.220687$\,  & $-\frac{a_0}{2\pi }0.216574$ \\
& & & & \\
 0.5 & $\frac{1}{2\pi }(2)$ & $\frac{a_0}{2\pi }1.82261$ & $\frac{a_0}{2\pi }0.285403$\,  & $-\frac{a_0}{2\pi }0.277177$ \\
& & & & \\
 0.6 & $\frac{1}{2\pi }(2)$ & $\frac{a_0}{2\pi }43.0296$\,  & $\frac{a_0}{2\pi }4.28244$\,  & $-\frac{a_0}{2\pi }4.27257$ \\
& & & & \\
 0.75 & $\frac{1}{2\pi }(1)$ & $\frac{a_0}{2\pi }0.821131$\,  & $\frac{a_0}{2\pi }0.0550438$\,  & $-\frac{a_0}{2\pi }0.0427044$ \\
& & & & \\
 1 & $\frac{1}{2\pi }(1)$ & $\frac{a_0}{2\pi }0.192793$ & $\frac{a_0}{2\pi }0.0254114$\,  & $-\frac{a_0}{2\pi }0.00895888$  \\
& & & & \\
\hline
\end{tabular}
\caption{Parameters of the effective action of Eq.\eqref{eq:Final_continuum_action} for several values of the anisotropy parameter $\lambda$.}
\label{table:coefficients}
\end{table}

\section{Spin-Spin correlation functions}
\label{app:Spin_Spin_expectation}

In this appendix we describe the procedure to obtain the spin-spin correlations shown in Sec \ref{sec:spin-correlations}. The first subsection below focuses specifically on the $S^zS^z$ correlations. The next subsection then deals with the $XY$ correlations. 

\subsection{$S^zS^z$ correlation functions}
\label{app:SzSz_expectation}

The fluctuating components of the $<S^z S^z>$ correlation can be expressed in terms of the magnetic field from the Jordan-Wigner transformation as shown below. This expectation can then be computed by introducing the usual source term $j_{\mu}$ coupled to a Chern-Simons term as follows 
\begin{equation}
	\begin{aligned}
	\langle S^z (x) S^z(y)\rangle \approx & \langle B(x)B(y) \rangle  \\ 
			\approx & - \theta ^2 \frac{1}{Z}\frac{\delta }{\delta  j_0(x)}\frac{\delta }{\delta _0 j(y)}\int  D A_{\mu } e^{i \frac{1}{2}\int d^3x d^3y A_{\mu }(x)G^{-1,\mu \nu }(x,y)A_{\lambda }(y)+i \int d^3x j_{\mu }(x)\epsilon ^{\mu \nu \lambda }\partial _{\nu }A_{\lambda }(x)}
	\end{aligned}
	\label{eq:SzSz_exp}
\end{equation}
where $G^{\mu\nu}$ is the continuum Green's function of the Lagrangian in Eq.\eqref{eq:Final_action_app} and $Z$ is the partition function. 

The $A_{\mu}$ fields can now be integrated out using the standard procedure  of shifting the fields $A_{\mu} \to a_{\mu} + \eta_{\mu}$ and then requiring that the terms linear in $\eta_{\mu}$ cancel. This gives the condition
\begin{equation}
	a_{\delta }(x) =-\int _{x'}G_{\delta \mu }(x,x') \epsilon ^{\mu \nu \lambda } \partial '_{\nu }j_{\lambda }(x')
\end{equation}
Substituting the above expression back into Eq.\eqref{eq:SzSz_exp} gives
\begin{equation}
	\begin{aligned}
		\langle S^z (x) S^z(y)\rangle \approx & - \theta ^2 \frac{\delta }{\delta  j_0(x)}\frac{\delta }{\delta _0 j(y)} e^{i\int d^3x d^3x'\; \left\{-\frac{1}{2} j_{\mu }(x) \left[\epsilon ^{\mu \nu \lambda }\epsilon ^{\delta \rho \epsilon }\partial _{\nu }\partial '_{\rho } G_{\lambda \epsilon} (x,x')\right]j_{\delta }( x') \right\}} \\
		= & i \theta^2 \epsilon^{0\nu\lambda} \epsilon^{0\rho\epsilon} \partial _{\nu }\partial^y_{\rho } G_{\lambda \epsilon}(x,y) 
	\end{aligned}
\end{equation}
The Green's function can be computed in the momentum space. Hence, the Fourier transform of the $S^zS^z$ spin correlation $f^{zz}(p)$ can now be expressed as
\begin{equation}
	f^{zz}(p) = \theta^2 \left(p_1^2 G_{22}(p) -p_1 p_2 G_{12}(p)-p_1 p_2 G_{21}(p)+p_2^2 G_{11}(p)\right)
\end{equation}
The propagator can be computed by introducing a gauge fixing term $\frac{\alpha}{2} (\partial^{\mu}A_{\mu})^2$
\begin{equation}
G^{-1}_{\mu\nu}(p)=\left(
\begin{array}{ccc}
 \alpha  \omega ^2+p_1^2 \epsilon +p_2^2 \epsilon  & p_1 \omega  (\epsilon -\alpha )+i \bar{\theta}  p_2 & p_2 \omega  (\epsilon -\alpha )-i \bar{\theta}  p_1 \\
 p_1 \omega  (\epsilon -\alpha )-i \bar{\theta}  p_2 & \alpha  p_1^2 - p_2^2 \chi +\omega ^2 \epsilon  & + p_1 p_2 (\chi + \alpha )-i \bar{\theta}  \omega  \\
 p_2 \omega  (\epsilon -\alpha )+i \bar{\theta}  p_1 & + p_1 p_2 (\chi + \alpha )+i \bar{\theta}  \omega  & \alpha  p_2^2-p_1^2 \chi +\omega ^2 \epsilon  \\
\end{array}
\right)
	\label{eq:Cont_Green_Func}
\end{equation}
where $\bar{\theta} = \theta + \theta_F$. The limit $\alpha \to \infty$ corresponds to the Lorentz gauge ($\partial^{\mu}A_{\mu} = 0$). In this gauge the above correlation yields
\begin{equation}
f^{zz}(p) \approx \theta^2 \frac{\epsilon \vec{p}^2}{ \epsilon^2 \omega^2 - \epsilon \chi \vec{p}^2 - (\theta + \theta_{F})^2}
\end{equation}
where $\vec{p}^2 = p_1^2 + p_2^2$.

Note that in the $XY$ regime $\theta + \theta_F = \frac{1}{\pi}$ for the case of $\frac{1}{3}$ filling on the Kagome lattice making the collective modes massive.
\begin{equation}
f^{zz}(p) \approx \theta^2 \frac{\epsilon \vec{p}^2}{ \epsilon^2 \omega^2 - \epsilon \chi \vec{p}^2 - (\frac{1}{\pi})^2}
\end{equation}
If the same calculation were performed on the square lattice (which is unfrustrated), the two terms would cancel (since $\theta + \theta_F = 0$) as observed in the main text and the collective modes are massless corresponding to Goldstone modes in the $XY$ regime.

\subsection{$XY$ correlation functions}
\label{app:XY_correlations}

Using the Jordan-Wigner transformation, the spin-spin correlation function $<S^+(x)S^-(y)>$ can be written as
\begin{equation}
	\begin{aligned}
		S^{XY}_{\Gamma}(x,y) = & \langle S^+(x)S^-(y)\rangle \\
			 \approx & \langle\psi^{\dagger}(x) e^{i \int_{\Gamma(x,y)} A_{\mu}dx^{\mu}} \psi(y) \rangle  \\
			= & \langle G_F(x,y;A_{\mu}) e^{i \int_{\Gamma(x,y)} A_{\mu}dx^{\mu}} \rangle_{A_{\mu}}
	\label{eq:XY_expectation}
	\end{aligned}
\end{equation}
where $G_F(x,y; A_{\mu})$ is the fermion Green function in the presence of the statistical gauge field $A_{\mu}$ and is obtained by integrating out the fermionic degrees of freedom. The average in the last step is over the statistical gauge fields $A_{\mu}$. Computing this expectation is involving due to the presence of the non-local string along the path $\Gamma(x,y)$ that starts from $x$ and ends at $y$. To simplify the above expression, the fermionic propagator can be expressed in terms of a Feynman path integral as a sum over histories of the particles as follows
\begin{equation}
	G_F(x,y; A_{\mu}) = -i \int_0^{\infty} dT \int D \vec{z}(t) e^{i S[\vec{z}(t)]}
\end{equation}
where the action $S$ is the action for non-relativistic particles coupled to the statistical gauge field
\begin{equation}
	S = \int_0^T dt \left\{ \frac{1}{2} \left( \frac{d\vec{z}}{dt} \right)^2 + \frac{dz^{\mu}}{dt}A_{\mu}(\vec{z}) \right\}
	\label{eq:B10}
\end{equation}
subject to the boundary conditions for a particle traveling from $y$ to $x$
\begin{equation}
	\begin{matrix}
		\lim_{t \to 0} \vec{z}(t) = \vec{y} & & & & \lim_{t \to T} \vec{z}(t) = \vec{x}
	\end{matrix}
\end{equation}
Note that the second term in the  action of Eq.\eqref{eq:B10} corresponds to another Wilson line but now traveling from point $y$ to point $x$. This combined with the Wilson line in Eq.\eqref{eq:XY_expectation} now creates a closed loop which we will call $\gamma$. For a problem with an energy gap and in the long-distance limit, the dominant trajectories are close to the classical trajectories. In this approximation the average over the different trajectories $\vec{z}(t)$ can be pulled outside of the integral for averaging over the statistical gauge fields. Hence, Eq.\eqref{eq:XY_expectation} can now be written as
\begin{equation}
	\begin{aligned}
		S^{XY}_{\Gamma}(x,y) = & \int_0^{\infty} dT \int D\vec{z}(t) e^{ i \int_0^T dt  \frac{1}{2} \left( \frac{d\vec{z}}{dt} \right)^2} \left< e^{i \int_{\gamma} A^{\mu}dz_{\mu}} \right>_{A_{\mu}} \\ 
			\approx & \sum_{\gamma} (Amp)_{\gamma} \left< e^{i \int_{\gamma} A^{\mu}dz_{\mu}} \right>_{A_{\mu}}	
	\end{aligned}
\end{equation}
where $(Amp)_{\gamma}$ is the amplitude associated with a path $\gamma$ and the set of closed curves $\{\gamma\}$ are the oriented sum of paths $\Gamma$ and the histories of the particle. 

Now, the computation of the Wilson loop expectation value can be performed by introducing a source term $J_{\mu}$ as follows
\begin{equation}
		\left< e^{i \int_{\gamma} A^{\mu}dz_{\mu}} \right>_{A_{\mu}}	= \left< e^{i \int d^3 z J_{\mu}(z)A^{\mu}(z)} \right>_{A_{\mu}}	
\end{equation}
where
\begin{equation}
J_{\mu}(z) = \left\{
	\begin{array}{lll}
		S_{\mu}(z)  & & \mbox{if } z \epsilon \gamma \\
		0 & & \textrm{otherwise}
	\end{array}
\right.
\end{equation}
where $S_{\mu}(z)$ is a unit vector tangent to the path $\gamma$ at $z$. In this form the above expectation can be written as
\begin{equation}
		\left< e^{i \int_{\gamma} A^{\mu}dz_{\mu}} \right>_{A_{\mu}}	= e^{-\frac{i}{2} \int_{x,y} J_{\mu}(x)G^{\mu\nu}(x,y)J_{\nu}(y) }	
\end{equation}
where $G^{\mu\nu}$ is the Green's function in the continuum for the statistical gauge fields shown above in Eq.\eqref{eq:Cont_Green_Func}. The exponent in the above integral has two contributions, one from the Maxwell like terms ($I_1$) and the other from the Chern-Simons like terms $I_2$. First, the $I_1$ term can be simplified to give
\begin{equation}
	I_1 = -\frac{\epsilon}{2} \int_{x,y} J_{\mu}(x) G_0(x,y; \bar{\theta}^2) J^{\mu}(y)
\end{equation}
where the propagator $G_0(x,y; \theta^2)$ can be approximated in the long-distance limit as follows
\begin{equation}
	\begin{aligned}
	G_0(x,y; \bar{\theta}^2) = \bra{x} \frac{1}{\epsilon^2 \partial^2 + \epsilon (\chi - \epsilon)\partial_i^2-\bar{\theta}^2} \ket{y}  
		\approx  -\frac{1}{\bar{\theta}^2} \delta(x-y)
	\end{aligned}
\end{equation}
Note here that we are assuming that the above propagator is massive (i.e $\bar{\theta} \neq 0$) as in the case of $\frac{1}{3}$ filling in the $XY$ limit of the Kagome lattice. This argument would clearly fail at other fillings or for instance in the case of the square lattice when $\bar{\theta} = 0$ changing the physics all together.

Hence, the integral $I_1$ can be approximated as
\begin{equation}
	\begin{aligned}
		I_1 \approx  \frac{\epsilon}{2 \bar{\theta}^2} \int_x J_{\mu}(x)J^{\mu}(x) 
			= -\frac{\epsilon}{2 (\theta+\theta_F)^2} L(\gamma)
	\end{aligned}
\end{equation}
where $L(\gamma)$ is the length of the path $\gamma$. The second integral due to the Chern-Simons term approximates to give (again in the long-distance limit)
\begin{equation}
	\begin{aligned}
		I_2 = & \frac{i \bar{\theta}}{2} \int_{x,y} J_{\mu}(x) \bra{x} \frac{1}{\partial^2 \left(\epsilon^2 \partial^2 + \epsilon (\chi - \epsilon)\partial_i^2-\bar{\theta}^2 \right) } \epsilon^{\mu\nu\lambda} \partial_{\lambda} \ket{y} J_{\nu}(y)\\
		\approx &  \frac{i}{2 \bar{\theta} } \int_{x,y} J_{\mu}(x)\epsilon^{\mu\nu\lambda} \bra{x} \frac{1}{\partial^2  } \ket{y} \partial_{\nu}J_{\lambda}(y) \\
		= &  \frac{i}{2\bar{\theta}} \int_{x,y} J_{\mu}(x)\epsilon^{\mu\nu\lambda} G_0(x,y;0) \partial_{\nu}J_{\lambda}(y)
	\end{aligned}
\end{equation}
The current $J_{\mu}(x)$ can be regarded as an electric current. With this interpretation the current can be related to a magnetic field $B_{\mu}(x)$ as follows
\begin{equation}
	B_{\mu}(x) = \int_y G_0(x,y;0) \epsilon_{\mu\nu\lambda}\partial^{\nu} J^{\lambda}(y)
\end{equation}
Now the second integral $I_2$ can be written in terms of a magnetic field as
\begin{equation}
	\begin{aligned}
		I_2 \approx \frac{i}{2\bar{\theta}} \frac{\Phi_{\gamma}}{\phi_0} = \frac{i}{2} \Phi_{\gamma}
	\end{aligned}
\end{equation}
where again at the semi-classical level we have approximated the field by the average flux $\Phi_{\gamma}$ over the path $\gamma$. This makes the above integral take the form of an Aharonov-Bohm phase over the path $\gamma$. In the last step we have used the fact that $\bar{\theta} = \theta + \theta_F = \frac{1}{\pi}$ and $\phi_0 = 2\pi$ (in units of $h = e = c= 1$).
Hence, the correlation can be approximated as
\begin{equation}
	S^{XY}_{\Gamma}(x,y) \approx \sum_{\gamma}(Amp)_{\gamma} e^{I_1 + I_2}
\end{equation}
where $I_1$ decays exponentially as the length of the path $L(\gamma)$ increases and $I_2$ is an Aharonov-Bohm phase associated with the path $\gamma$. The Aharonov-Bohm term in the above expression would depend 
on the two points $x$ and $y$ and the area 
enclosed by the path $\gamma$. In the main text we have considered two possible situations, one where the area of the path $\gamma$ is zero and doesn't lead to any Aharonov-Bohm phase and the other where there are several different Aharonov-Bohm phases that can arise. 
\end{widetext}


\end{document}